\documentclass[3p]{elsarticle}

\usepackage{amsmath}
\usepackage{amsthm}
\usepackage{amssymb}
\usepackage{graphicx}
\usepackage{esint}
\usepackage{xcolor}

\allowdisplaybreaks

\usepackage{lineno}
\PassOptionsToPackage{hyphens}{url}
\usepackage[pdfencoding=auto,psdextra,pdfauthor=author]{hyperref}
\usepackage{bookmark}

\usepackage{tikz}
\usepackage{chemfig}
\usepackage{mol2chemfig}
\usepackage[version=3]{mhchem}

\usepackage{chngcntr}

\biboptions{numbers,sort&compress}

\modulolinenumbers[1]

\journal{Chaos, Solitons \& Fractals}

\begin{document}

\begin{frontmatter}

\title{Thermal stability of solitons in protein \texorpdfstring{$\alpha$}{alpha}-helices}

\author[address1]{Danko D. Georgiev\corref{mycorrespondingauthor}}
\ead{danko.georgiev@mail.bg}
\cortext[mycorrespondingauthor]{Corresponding author}

\author[address2]{James F. Glazebrook}
\ead{jfglazebrook@eiu.edu}

\address[address1]{Institute for Advanced Study, 30 Vasilaki Papadopulu Str., Varna 9010, Bulgaria}
\address[address2]{Department of Mathematics, Eastern Illinois University, 600 Lincoln Ave., Charleston, Illinois 61920-3099, USA}

\begin{abstract}
Protein $\alpha$-helices provide an ordered biological environment that is conducive to soliton-assisted energy transport. The nonlinear interaction between amide~I excitons and phonon deformations induced in the hydrogen-bonded lattice of peptide groups leads to self-trapping of the amide~I energy, thereby creating a localized quasiparticle (soliton) that persists at zero temperature. The presence of thermal noise, however, could destabilize the protein soliton and dissipate its energy within a finite lifetime. In this work, we have computationally solved the system of stochastic differential equations that govern the quantum dynamics of protein solitons at physiological temperature, $T=310$~K, for either a single isolated $\alpha$-helix spine of hydrogen bonded peptide groups or the full protein $\alpha$-helix comprised of three parallel $\alpha$-helix spines. The simulated stochastic dynamics revealed that although the thermal noise is detrimental for the single isolated $\alpha$-helix spine, the cooperative action of three amide~I exciton quanta in the full protein $\alpha$-helix ensures soliton lifetime of over $30$~ps, during which the amide~I energy could be transported along the entire extent of an 18-nm-long $\alpha$-helix. Thus, macromolecular protein complexes, which are built up of protein $\alpha$-helices could harness soliton-assisted energy transport at physiological temperature. Because the hydrolysis of a single adenosine triphosphate molecule is able to initiate three amide~I exciton quanta, it is feasible that multiquantal protein solitons subserve a variety of specialized physiological functions in living systems.
\end{abstract}

\begin{keyword}
Davydov soliton\sep Langevin dynamics\sep protein $\alpha$-helix\sep soliton lifetime\sep thermal noise
\end{keyword}

\date{November 16, 2021}

\end{frontmatter}

%\linenumbers

\section*{Highlights}

\begin{itemize}

\item Secondary structure of protein $\alpha$-helices supports the generation of moving Davydov solitons.

\item Lateral coupling between protein $\alpha$-helix spines is indispensable for soliton thermal stability.

\item Lateral coupling between spines is mimicked by tripled mass of the phonon lattice unit cells.

\item Soliton lifetime of 30--40~ps ensures energy transport in proteins at physiological temperature.

\item Langevin dynamics of the protein lattice leads to stochastic trajectories of Davydov solitons.

\end{itemize}

\pagebreak

\section{Introduction}

Proteins are biological macromolecules that sustain life through execution
of main physiological functions such as transportation, contraction,
signal transduction or catalysis of biochemical reactions \cite{McLachlan1972,Ouzounis2003,Rodwell2018}.
To achieve their diverse functions, proteins rely on efficient utilization
of energy quanta released by the hydrolysis of adenosine triphosphate
(ATP) to adenosine diphosphate (ADP) and inorganic phosphate (Pi).
The Gibbs free energy of ATP hydrolysis (measured in eV) depends on the chemical concentration of substrates and products \cite{Barclay2020,Kammermeier1982,Weiss2005}
\begin{equation}
\Delta G_{\textrm{ATP}}=\Delta G_{\textrm{ATP}}^{0}+\frac{k_{B}T}{q_{e}}\ln\left(\frac{[\textrm{ADP}][\textrm{Pi}]}{[\textrm{ATP}]}\right)\label{eq:ATP}
\end{equation}
where $k_{B}=1.380649\times10^{-23}$~J/K is the Boltzmann constant
\cite{Newell2019}, $T=310$~K is the physiological temperature of
the human body, $q_{e}=1.602176634\times10^{-19}$~C is the elementary
electric charge \cite{Newell2019}, and $\Delta G_{\textrm{ATP}}^{0}=-0.31611$~eV
is the standard Gibbs free energy of ATP hydrolysis measured under
standard conditions in which the concentrations of all chemicals in
the reaction, $[\textrm{ATP}]$, $[\textrm{ADP}]$, and $[\textrm{Pi}]$,
are 1~molar (M) \cite{Berg2002,Cooper2019}. For physiological intracellular
conditions with $[\textrm{ATP}]=6.18$~mM, $[\textrm{ADP}]=0.05$~mM
and $[\textrm{Pi}]=0.88$~mM \cite{Kammermeier1982}, the free
energy released by ATP hydrolysis computed from \eqref{eq:ATP} is
about $0.63$~eV. This amount of energy is sufficient to excite three
amide~I exciton (C=O stretching) quanta, which couple strongly with the lattice of hydrogen
bonds that support the secondary structure of protein $\alpha$-helices
\cite{Pauling1951}.

\begin{figure*}[t]
\begin{centering}
\includegraphics[width=130mm]{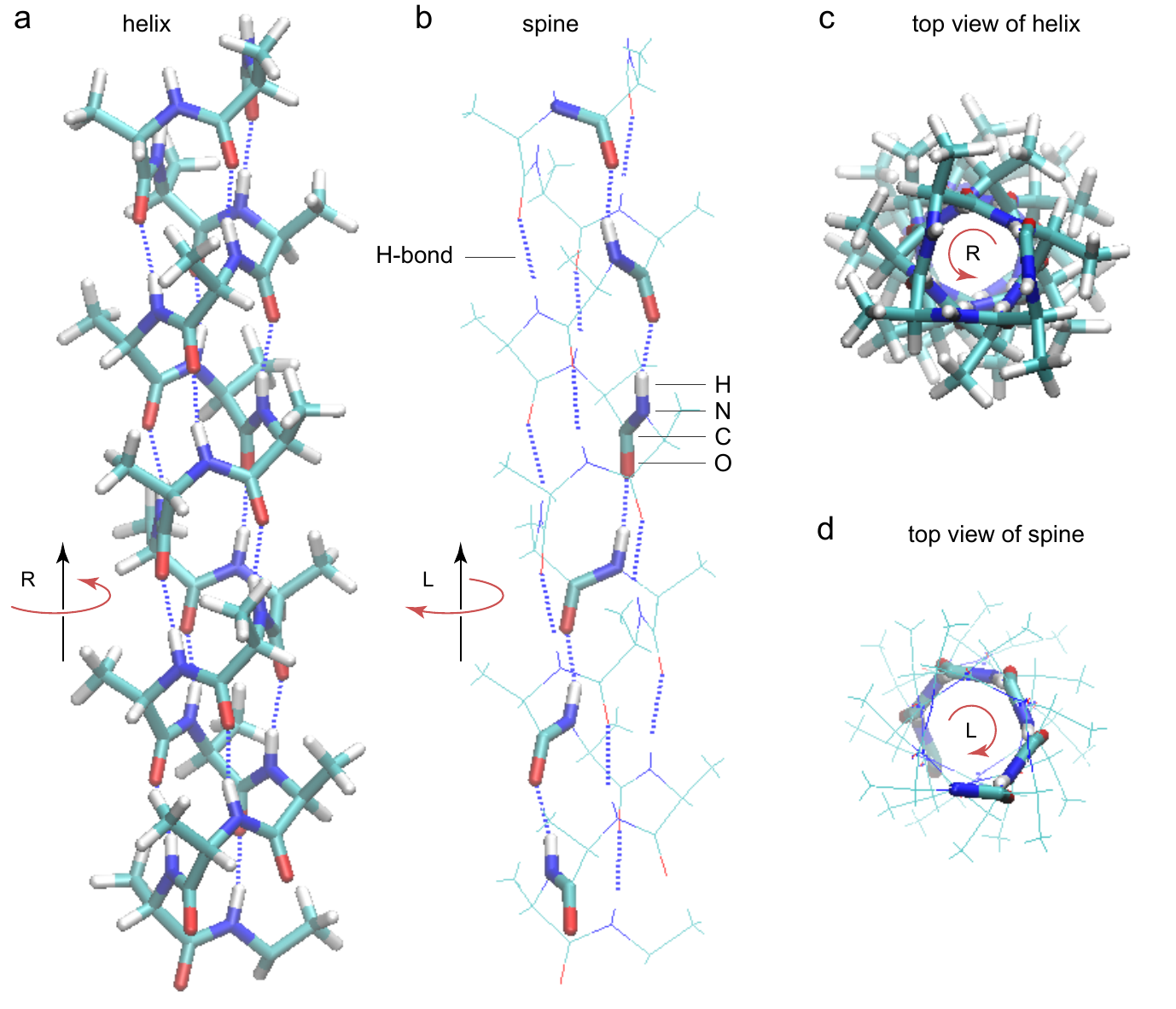}
\par\end{centering}

\caption{\label{fig:1}Three-dimensional structure of a poly-alanine protein
$\alpha$-helix stabilized by three chains of hydrogen bonds (dashed
lines) referred to as $\alpha$-helix spines. (a) Side view of the
protein $\alpha$-helix. (b) Side view of a single protein $\alpha$-helix
spine. (c) Top view of the protein $\alpha$-helix. (d) Top view of
a single protein $\alpha$-helix spine. H~atoms are shown in white,
N~atoms in blue, C~atoms in green, and O~atoms in red. The rise of
right-handed (R) spiral occurs with counter-clockwise rotation, whereas
the rise of left-handed (L) spiral occurs with clockwise rotation.}
\end{figure*}

Structurally, the protein $\alpha$-helix is a right-handed spiral
that contains 3.6~amino acid residues for each full turn (Fig.~\ref{fig:1}a).
Three parallel chains of hydrogen-bonded peptide groups $\cdots$H-N-C=O$\cdots$H-N-C=O$\cdots$,
referred to as $\alpha$-helix spines, stabilize the spiral (Fig.~\ref{fig:1}b).
The amino acid residues in the protein $\alpha$-helix
are tightly packed with an inner diameter of about 300~pm, which makes
the $\alpha$-helix inaccessible for entry of water molecules from the solvent (Fig.~\ref{fig:1}c).
Each of the three individual $\alpha$-helix spines
also rotates, but in the opposite direction, to form a left-handed
spiral (Fig.~\ref{fig:1}d).

The transportation of metabolic energy along protein \mbox{$\alpha$-helices},
as described by Davydov's theory of interacting amide~I excitations
(excitons) and distortions in the lattice of hydrogen bonds (phonons),
is modeled by the following generalized Hamiltonian \cite{Davydov1976,Davydov1979,Davydov1981}
\begin{equation}
\hat{H}=\hat{H}_{\textrm{ex}}+\hat{H}_{\textrm{ph}}+\hat{H}_{\textrm{int}}\label{eq:H}
\end{equation}
with
\begin{align}
\hat{H}_{\textrm{ex}} & =\sum_{n}\left[E_{0}\hat{a}_{n}^{\dagger}\hat{a}_{n}-J\left(\hat{a}_{n}^{\dagger}\hat{a}_{n+1}+\hat{a}_{n}^{\dagger}\hat{a}_{n-1}\right)\right]\\
\hat{H}_{\textrm{ph}} & =\frac{1}{2}\sum_{n}\left[\frac{\hat{p}_{n}^{2}}{\tilde{M}}+\tilde{w}\left(\hat{u}_{n+1}-\hat{u}_{n}\right)^{2}\right]\\
\hat{H}_{\textrm{int}} & =\sum_{n}\left[\chi_{r}\hat{u}_{n+1}+\left(\chi_{l}-\chi_{r}\right)\hat{u}_{n}-\chi_{l}\hat{u}_{n-1}\right]\hat{a}_{n}^{\dagger}\hat{a}_{n}
\end{align}
where $E_{0}=0.2$~eV is the energy of the amide~I exciton, $J=967.4$~$\mu$eV
is the dipole--dipole coupling energy between neighboring
amide~I oscillators, $\hat{a}_{n}^{\dagger}$ and $\hat{a}_{n}$ are
the exciton creation and annihilation operators satisfying the commutation
relations $[a_{i},a_{j}^{\dagger}]\equiv a_{i}a_{j}^{\dagger}-a_{j}^{\dagger}a_{i}=\delta_{ij}$
and $[a_{i}^{\dagger},a_{j}^{\dagger}]=[a_{i},a_{j}]=0$, $\hat{p}_{n}$~is
the momentum operator and $\hat{u}_{n}$ is the displacement operator
from the equilibrium position of the lattice site~$n$, $\tilde{M}$
is the mass of each lattice site indexed by $n$, $\tilde{w}$ is
the spring constant of the hydrogen bonds in the lattice \cite{Davydov1976,Davydov1979,Davydov1981,Scott1992},
$\chi_{r}$ and $\chi_{l}$ are anharmonic parameters arising from
the coupling between the amide~I exciton and the phonon lattice
displacements, respectively, to the right or to the left, \cite{Luo2017,Georgiev2019a,Georgiev2019b,Georgiev2020a,Georgiev2020b}.

The model Hamiltonian \eqref{eq:H} is traditionally used to study
the quantum dynamics of solitons propagating along a single $\alpha$-helix
spine, where the individual lattice sites $n$ are taken to be peptide
groups with mass $M=1.9\times10^{-25}$~kg and the spring constant
of the hydrogen bonds between neighboring peptide groups is $w=13$~N/m \cite{Scott1982,Cruzeiro1988}.
However, A.~C.~Scott recognized that
the same Hamiltonian \eqref{eq:H} also describes the propagation
of symmetric solitons in the protein $\alpha$-helix if the individual
lattice sites~$n$ are taken to be protein unit cells comprised
from three consecutive peptide groups in the protein primary structure
with total mass $\tilde{M}=3M=5.7\times10^{-25}$~kg and the combined
spring constant of the three hydrogen bonds between neighboring peptide
unit cells is set to $\tilde{w}=3w=39$ N/m \cite{Scott1992}.
Here, we use the latter fact in order to study the effects of thermal noise on
the soliton lifetime, and investigate the potential unexpected consequences
of a theoretic consideration of a single $\alpha$-helix
spine isolated outside of the stable structure of the massive protein
$\alpha$-helix. In particular, we show that the single isolated $\alpha$-helix
spine is a poor model of protein $\alpha$-helix as it lacks thermal stability.
We further strengthen our analysis by considering computer simulations of the more elaborate, three-spine model of protein
$\alpha$-helix.

The organization of this work is as follows: In Section~\ref{sec:2}, we present a system of Langevin equations that model the quantum dynamics of protein $\alpha$-helix in the presence of physiological thermal noise at $T=310$~K. In Section~\ref{sec:3}, we provide numeric values of various biophysical parameters, together with the initial conditions for the amide~I exciton pulse and the phonon lattice, which are required for computer simulations. Next, in Section~\ref{sec:4} we compare the thermal stability of a single isolated $\alpha$-helix spine and the massive single-chain model of protein $\alpha$-helix. Finally, in Section~\ref{sec:5} we investigate the full three-spine model of protein $\alpha$-helix and demonstrate the importance of the lateral coupling between the spines for ensuring thermal stability of the generated protein solitons for 30-40 ps, which is sufficient for the amide~I energy to be transported along the entire extent of an 18-nm-long $\alpha$-helix.

\section{Equations of motion with thermal noise}
\label{sec:2}

Quantum equations of motion for multiquantal states of amide~I energy
can be derived from the Schr\"{o}dinger equation
\begin{equation}
\imath\hbar\frac{d}{dt}|\Psi\rangle=\hat{H}|\Psi\rangle\label{eq:schrodinger}
\end{equation}
using a generalized ansatz state vector that approximates the exact
solution in the form \cite{Zolotaryuk1988,Kerr1990}
\begin{equation}
|\Psi(t)\rangle=|\psi_{\textrm{ex}}(t)\rangle|\psi_{\textrm{ph}}(t)\rangle \label{eq:ansatz}
\end{equation}
with
\begin{align}
|\psi_{\textrm{ex}}(t)\rangle & =\frac{1}{\sqrt{Q!}}\left[\sum_{n}a_{n}(t)\hat{a}_{n}^{\dagger}\right]^{Q}|0_{\textrm{ex}}\rangle \label{eq:Hartree}\\
|\psi_{\textrm{ph}}(t)\rangle & =e^{-\frac{\imath}{\hbar}\sum_{j}\left(b_{j}(t)\hat{p}_{j}-c_{j}(t)\hat{u}_{j}\right)}|0_{\textrm{ph}}\rangle \label{eq:coherent}
\end{align}
where $|0_{\textrm{ex}}\rangle$~is the vacuum state of amide~I excitons,
$|0_{\textrm{ph}}\rangle$~is the vacuum state of phonons,
and we identify the following two expectation values:
$b_{j}=\langle\Psi(t)|\hat{u}_{j}|\Psi(t)\rangle$
and $c_{j}=\langle\Psi(t)|\hat{p}_{j}|\Psi(t)\rangle$ \cite{Georgiev2020a}.

Using the generalized Ehrenfest theorem for the time dynamics of the
expectation values of the latter two observables gives
\begin{align}
\imath\hbar\frac{d}{dt}b_{j} & =\langle\Psi(t)|\left[\hat{u}_{j},\hat{H}\right]|\Psi(t)\rangle\label{eq:Ehrenfest-1}\\
\imath\hbar\frac{d}{dt}c_{j} & =\langle\Psi(t)|\left[\hat{p}_{j},\hat{H}\right]|\Psi(t)\rangle\label{eq:Ehrenfest-2}
\end{align}
From \eqref{eq:Ehrenfest-1} follows that $c_j = \tilde{M} \frac{d}{dt}b_{j}$ \cite{Georgiev2019a}, which allows us to express the initial lattice state only in terms of $b_j(0)$ and $\frac{d}{dt}b_{j}(0)$ as follows
\begin{equation}
|\psi_{\textrm{ph}}(0)\rangle=e^{-\frac{\imath}{\hbar}\sum_{j}\left(b_{j}(0)\hat{p}_{j}-\tilde{M}\left[\frac{d}{dt}b_{j}(0)\right]\hat{u}_{j}\right)}|0_{\textrm{ph}}\rangle
\end{equation}
Further consideration of the expectation value
$\langle\Psi(t)|\hat{a}_{n}^{\dagger}\hat{a}_{n}|\Psi(t)\rangle=Q\,|a_{n}|^{2}$
together with the inner product based on the Schr\"{o}dinger equation
\begin{equation}
\frac{1}{\sqrt{Q!}}\langle\psi_{\textrm{ph}}(t)|\langle0_{\textrm{ex}}|\left(\hat{a}_{n}\right)^{Q}\imath\hbar\frac{d}{dt}|\Psi(t)\rangle=\frac{1}{\sqrt{Q!}}\langle\psi_{\textrm{ph}}(t)|\langle0_{\textrm{ex}}|\left(\hat{a}_{n}\right)^{Q}\hat{H}|\Psi(t)\rangle
\end{equation}
after straightforward, but laborious calculations (for a complete
step-by-step derivations see \cite{Georgiev2020a,Kerr1990,Kerr1987}),
results in the following system of gauge transformed quantum equations
of motion
\begin{eqnarray}
\imath\hbar\frac{d}{dt}a_{n} & = & -J\left(a_{n+1}+a_{n-1}\right)+\left[\chi_{r}b_{n+1}+(\chi_{l}-\chi_{r})b_{n}-\chi_{l}b_{n-1}\right]a_{n}\label{eq:gauge-1}\\
\tilde{M}\frac{d^{2}}{dt^{2}}b_{n} & = & \tilde{w}\Big(b_{n-1}-2b_{n}+b_{n+1})-Q\Big(\chi_{r}\left|a_{n-1}\right|^{2}+(\chi_{l}-\chi_{r})\left|a_{n}\right|^{2}-\chi_{l}\left|a_{n+1}\right|^{2}\Big)\label{eq:gauge-2}
\end{eqnarray}
The latter system of equations \eqref{eq:gauge-1} and \eqref{eq:gauge-2}
governs the quantum dynamics of protein solitons in the absence of
thermal noise ($T=0$~K). In order to introduce the effects of a thermal
bath coupled to the protein, one needs to modify the lattice dynamics
\eqref{eq:gauge-2} with the inclusion of a damping term $-\tilde{M}\Gamma\frac{d}{dt}b_{n}$
and a thermal noise term $\eta_{n}(t)$ that obey the fluctuation--dissipation
theorem \cite{Kubo1966}. The latter condition implies that the correlation
function for the random force is
\begin{equation}
\langle\eta_{n}(t)\eta_{n'}(t')\rangle=2\tilde{M}\Gamma k_{B}T\delta_{nn'}\delta(t-t')
\end{equation}
where $\delta_{nn'}$ is the Kronecker delta, $\delta(t-t')$ is the
Dirac delta function, and $\Gamma=0.005\sqrt{\frac{\tilde{w}}{\tilde{M}}}$
is the lowest (non-zero) frequency of the protein lattice used in
previous computational studies \cite{Halding1987,Lomdahl1990,Forner1991}.

In terms of the It\^{o} calculus \cite{Ikeda1996,Accardi2007,Schilling2012},
if we have a real-valued continuous-time stochastic Wiener process
$W_{n}(t)$ with zero drift $\langle W_{n}(t)\rangle=0$ and unit volatility $\langle[W_{n}(t)]^{2}\rangle=t$, then its formal derivative $\frac{d}{dt}W_{n}(t)$ is
a white noise process such that
\begin{align}
\left\langle\frac{d}{dt}W_{n}(t)\right\rangle & =0\\
\left\langle\left[\frac{d}{dt}W_{n}(t)\right]\left[\frac{d}{dt}W_{n}(t')\right]\right\rangle & =\delta\left(t-t'\right)
\end{align}
Therefore, the introduction of stochastic differentials \cite{Ito1975}
converts \eqref{eq:gauge-2} into a Langevin equation \cite{Langevin1908,Ford1987,Araujo2019,deOliveira2020}
\begin{equation}
\frac{d^{2}}{dt^{2}}b_{n}=\frac{\tilde{w}}{\tilde{M}}\Big(b_{n-1}-2b_{n}+b_{n+1})-\frac{Q}{\tilde{M}}\Big(\chi_{r}\left|a_{n-1}\right|^{2}+(\chi_{l}-\chi_{r})\left|a_{n}\right|^{2}-\chi_{l}\left|a_{n+1}\right|^{2}\Big)-\Gamma\frac{d}{dt}b_{n}+\sqrt{\frac{2\Gamma k_{B}T}{\tilde{M}}}\frac{d}{dt}W_{n}(t)\label{eq:SDE}
\end{equation}

\section{Computer simulations of stochastic soliton dynamics}
\label{sec:3}

The length of the simulated protein $\alpha$-helix in the present study was taken to be
$18$~nm, which corresponds to $n_{\max}=40$ lattice sites separated
by a distance $r=0.45$~nm between any two neighboring sites along the
main axis. Contrary to ordinary differential equations where the obtained
solutions are deterministic, computer simulations of stochastic processes
generate different observable paths for different simulation runs.
In order to be able to perform direct comparisons of the resulting
stochastic dynamics for different number~$Q$ of amide~I exciton quanta
or different $\tilde{M}$, we have randomly generated sets $S_{i}$
of $n_{\max}=40$ Wiener processes $\{W_{1}(t),W_{2}(t),\ldots,W_{40}(t)\}$
each of which is discretized in time steps of $\Delta t=5$ fs. Next,
we have thermalized the phonon lattice in the absence of exciton quanta,
$Q=0$, in order to extract initial $b_{n}(0)$ and $\frac{d}{dt}b_{n}(0)$
values of a protein lattice state $|\psi_{\textrm{ph}}(0)\rangle$ that has arrived exactly at temperature
of $T=310$~K, namely
\begin{equation}
\frac{1}{2}\sum_{n}\tilde{M}\left(\frac{db_{n}(0)}{dt}\right)^{2}=\frac{1}{2}N k_{B}T
\end{equation}
where $N$ is the total number of lattice sites.
Then, soliton simulations were performed using a discrete Gaussian
pulse of amide~I energy applied over five consecutive lattice sites~$n$ given by
\begin{equation}
\left\{ A_{2}e^{-\imath2\Delta\omega},A_{1}e^{-\imath\Delta\omega},A_{0},A_{1}e^{+\imath\Delta\omega},A_{2}e^{+\imath2\Delta\omega}\right\} \label{eq:pulse}
\end{equation}
where $A_{0}=\sqrt{0.322}$, $A_{1}=\sqrt{0.24}$, $A_{2}=\sqrt{0.099}$,
and the phase modulation introduced through the parameter $\Delta\omega=\frac{\pi}{12}$
ensures that the soliton has a positive initial momentum \cite{Georgiev2020b} given by
\begin{equation}
\langle\Psi(0)|\hbar\hat{k}|\Psi(0)\rangle=\frac{2\hbar}{r}\sin\left(\Delta\omega\right)\left(A_{0}A_{1}+A_{1}A_{2}\right)
\end{equation}
In the continuum approximation, Davydov and co-workers have shown that the solitons exhibit a characteristic sech-squared exciton probability distribution \cite{Davydov1976,Davydov1979}. For the utilization of energy by biological systems, however, it is advantageous if any sufficiently localized initial pulse of exciton energy is able to reshape into a soliton-like form with only minimal dispersion of energy into the so-called soliton `tails' \cite{Brizhik1983,Brizhik1993}. Consequently, in order to demonstrate that the protein solitons are easy to generate in living matter, in this present work we have chosen to apply general Gaussian exciton pulses, as we also did previously \cite{Georgiev2019a,Georgiev2019b,Georgiev2020a,Georgiev2020b}.
In~\ref{app-E}, we show that discrete sech-squared initial exciton pulses indeed exhibit dynamics that is very similar to the one obtained with discrete Gaussian exciton pulses.

To simulate solitons inside a single protein $\alpha$-helix spine,
we used $\tilde{M}=1M$ and $\tilde{w}=1w$, whereas for solitons
inside the massive single-chain model of a protein $\alpha$-helix, we used $\tilde{M}=3M$ and
$\tilde{w}=3w$. The exciton--phonon interaction was considered to
be isotropic with equal left and right coupling parameters, $\chi_{l}=\chi_{r}=35$
pN \cite{Scott1992,Georgiev2020a}. The present choice of modeling
parameters ensures direct comparability of the resulting simulations
with those reported in previous works \cite{Georgiev2019a,Georgiev2019b,Georgiev2020a,Georgiev2020b,Cruzeiro1988,Forner1991,Lomdahl1985}.

\section{Comparison between single \texorpdfstring{$\alpha$}{alpha}-helix spine and massive single-chain model of protein \texorpdfstring{$\alpha$}{alpha}-helix}
\label{sec:4}

\subsection{Quantum dynamics at zero temperature}

To set up the appropriate base case for assessing the influence of
thermal noise on the lifetime of protein solitons, we have first simulated
the quantum dynamics of solitons at zero temperature, $T=0$\,K,
in a single isolated protein $\alpha$-helix spine or in the massive single-chain model of a protein $\alpha$-helix.

\subsubsection{Single \texorpdfstring{$\alpha$}{alpha}-helix spine}

\begin{figure*}[t]
\begin{centering}
\includegraphics[width=140mm]{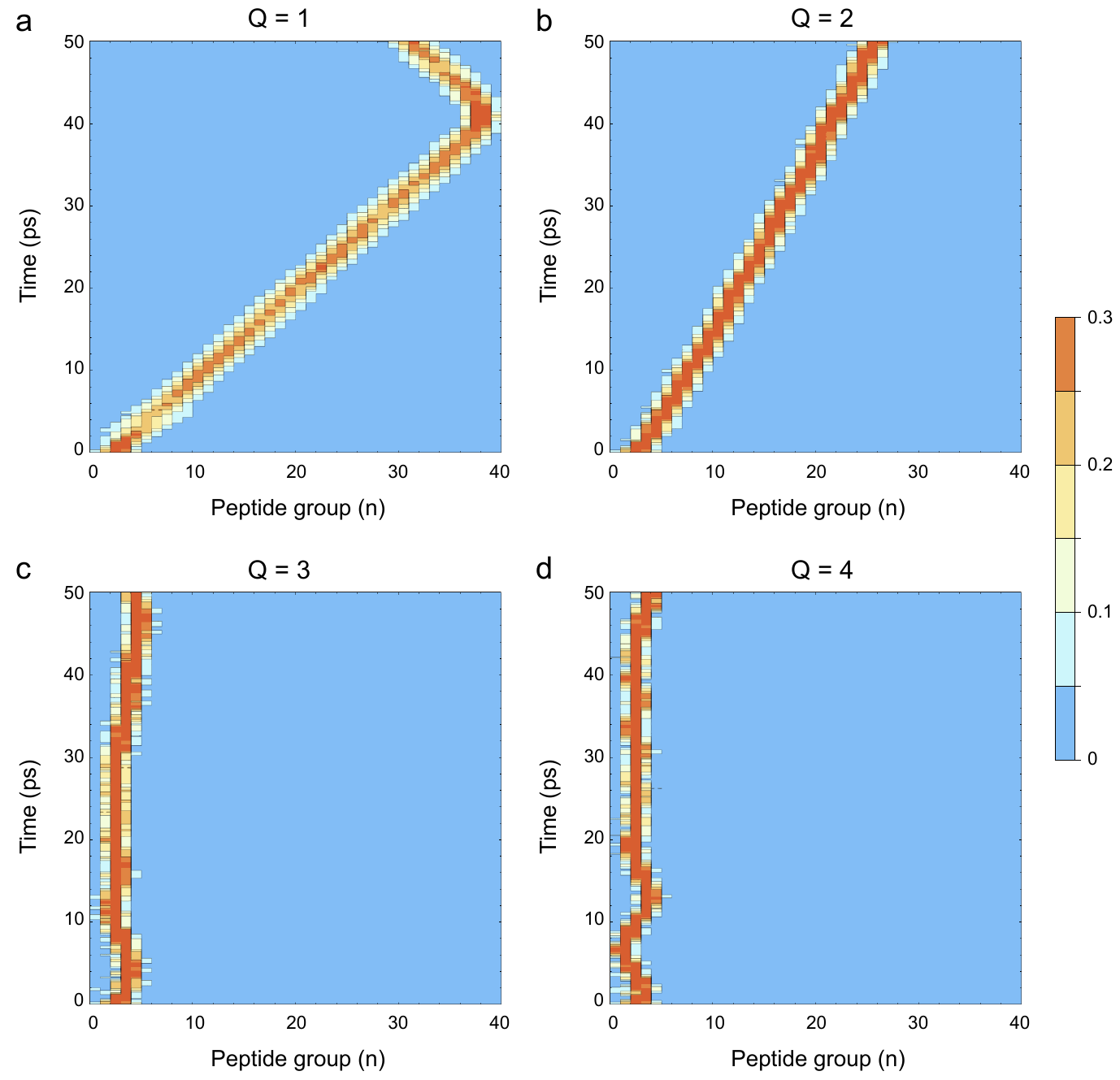}
\par\end{centering}

\caption{\label{fig:2}Dynamics of a soliton moving inside a single
$\alpha$-helix spine with $\tilde{M}=1M$ and $\tilde{w}=1w$ at
$T=0$~K, with different number~$Q$ of amide~I energy quanta visualized
through the exciton quantum probability $|a_{n}|^{2}$ at each peptide
group~$n$. The Gaussian pulse of amide~I energy is applied at the
N-end of a protein $\alpha$-helix spine with $n_{\max}=40$ peptide
groups. The lifetime of the protein soliton exceeds $50$~ps. For
$Q\geq3$, the soliton is pinned.}
\end{figure*}

In the model of a single $\alpha$-helix spine with $\tilde{M}=1M$
and $\tilde{w}=1w$, the actual number $Q$ of amide~I exciton quanta
was crucial for surpassing the upper threshold for which the self-trapping
of the soliton is too strong to allow soliton motion, thereby pinning
the soliton at the site of its creation (Fig.\,\ref{fig:2}). In
general, increasing $Q$ decreases the gauge transformed soliton energy
given by the initial exciton expectation value \cite{Georgiev2020b}
\begin{equation}
\langle\Psi(0)|\hat{H}_{\textrm{ex}}|\Psi(0)\rangle-QE_{0}  =-QJ\sum_{n}\left[a_{n}^{*}(0)a_{n+1}(0)+a_{n}(0)a_{n+1}^{*}(0)\right]
=-4QJ\left(A_{0}A_{1}+A_{1}A_{2}\right)\cos(\Delta\omega)
%\approx-1.67 QJ
\end{equation}
and the resulting stronger self-trapping effect slows down the speed
of the soliton.

For $Q=1$, the self-trapping is well above the lower threshold for
soliton formation and the soliton speed is 439 m/s (Fig.\,\ref{fig:2}a).
For $Q=2$, the stronger self-trapping slows down the soliton speed
to 234 m/s (Fig.~\ref{fig:2}b). For $Q\geq3$, the soliton remains
pinned at the site of its creation (Fig.~\ref{fig:2}c,d). These results
are consistent with previous reports that increasing the strength
of the nonlinear exciton--phonon interaction leads to pinning of the
solitons, which in turn makes them unsuitable for transport of energy
as they do not propagate at all \cite{Forner1990,Forner1991,Forner1991b,Forner1991c,Georgiev2019a,Georgiev2020a}.
Thus, the theoretical modeling of a single $\alpha$-helix spine extracted
outside of the full protein $\alpha$-helix already has some unintended
consequences upon the biological utility of solitons even in the absence
of any thermal noise. In particular, with respect to the number $Q$
of amide~I exciton quanta, the window with moving solitons is much
narrower for the single protein $\alpha$-helix spine as compared
with the full protein $\alpha$-helix, which we elaborate next.

\subsubsection{Massive single-chain model of protein \texorpdfstring{$\alpha$}{alpha}-helix}

\begin{figure*}[t]
\begin{centering}
\includegraphics[width=140mm]{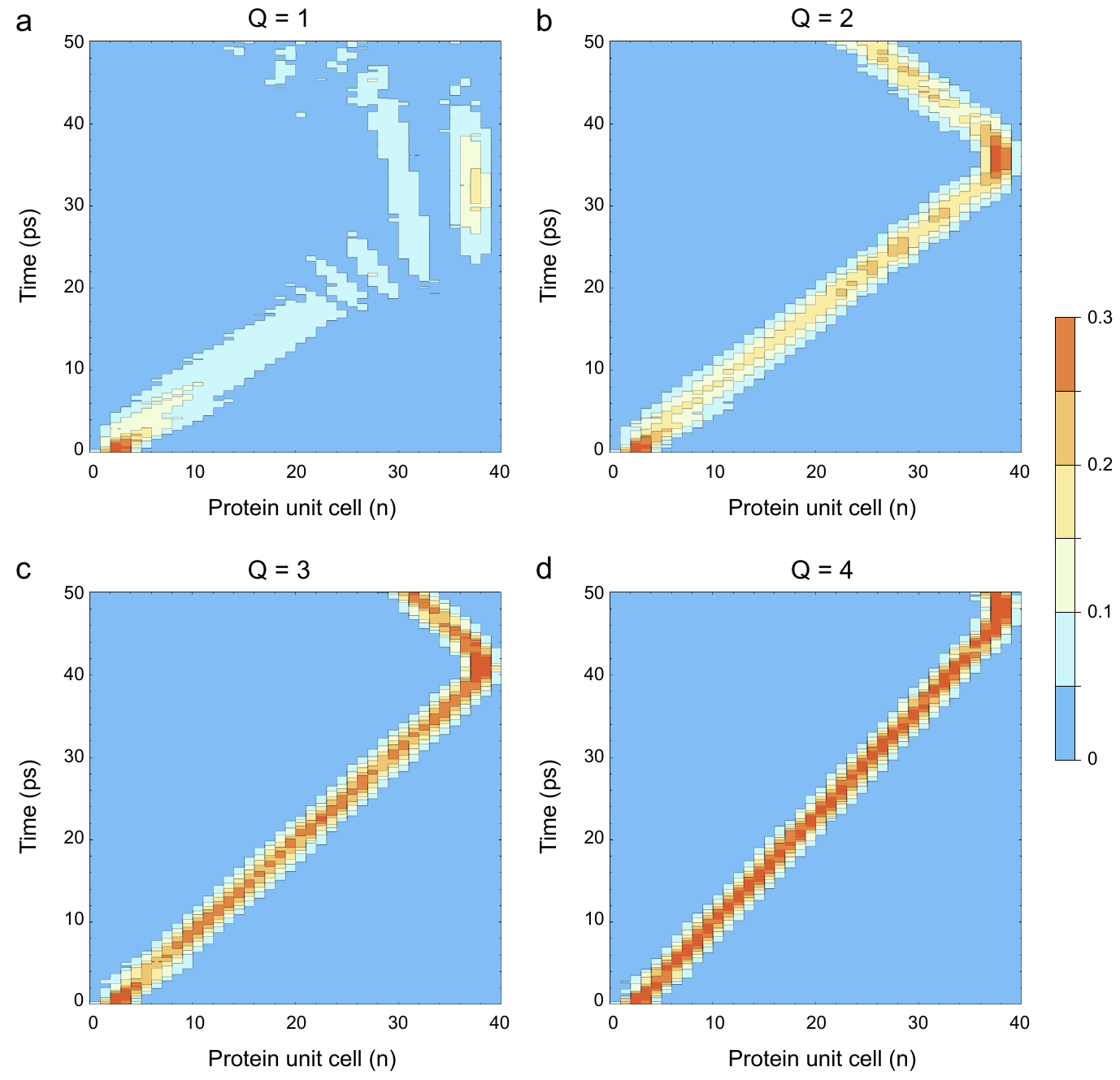}
\par\end{centering}

\caption{\label{fig:3}Dynamics of a soliton moving inside the massive single-chain model of a protein
$\alpha$-helix with $\tilde{M}=3M$ and $\tilde{w}=3w$ at $T=0$~K, with different number $Q$ of amide~I energy quanta visualized
through the exciton quantum probability $|a_{n}|^{2}$ at each protein
unit cell~$n$. The~Gaussian pulse of amide~I energy is applied at
the N-end of a protein $\alpha$-helix with $n_{\max}=40$ unit cells,
each of which is comprised of three peptide groups. For $Q\geq2$, the
lifetime of the protein soliton exceeds $50$~ps.}
\end{figure*}

In the massive single-chain model of a protein $\alpha$-helix with $\tilde{M}=3M$
and $\tilde{w}=3w$, the actual number~$Q$ of amide~I exciton quanta
was crucial for surpassing the lower threshold for which the self-trapping
is strong enough to prevent the soliton dispersal (Fig.~\ref{fig:3}).
In the presence of a single amide~I exciton quantum, $Q=1$, the nonlinear
interaction between the exciton and the distortion of the phonon lattice,
given by $Q\chi_{r}\left|a_{n-1}\right|^{2}$ and $Q\chi_{l}\left|a_{n+1}\right|^{2}$
terms, is not strong enough to prevent dispersion (Fig.~\ref{fig:3}a).
From the viewpoint of a complete model with three $\alpha$-helix
spines (see \ref{app-F}), the latter observation can be explained by the reduction of
the exciton quantum probability amplitudes~$a_{n}$ by $\frac{1}{\sqrt{3}}$ due to their distribution
over the three spines thereby decreasing the exciton--phonon nonlinear
interaction below the threshold for soliton formation. For $Q\geq2$,
the self-trapping effect is above the lower threshold for soliton
formation, but below the upper threshold for soliton pinning (Figs.
\ref{fig:3}b-d). For $Q=2$ the soliton propagates with speed of
500 m/s (Fig.~\ref{fig:3}b), for $Q=3$ the speed is decreased to
439 m/s (Fig.~\ref{fig:3}c), and for $Q=4$ the speed is 367 m/s
(Fig.~\ref{fig:3}d). In the absence of thermal noise, for $Q\geq2$
the soliton lifetime was confirmed to exceed with at least two orders of magnitude the
simulation period of 50~ps, despite possible destabilizing effects due to finite working precision and to numerical rounding-off in computer simulations.

\subsection{Quantum dynamics at physiological temperature}

\subsubsection{Single \texorpdfstring{$\alpha$}{alpha}-helix spine}

\begin{figure*}[t]
\begin{centering}
\includegraphics[width=140mm]{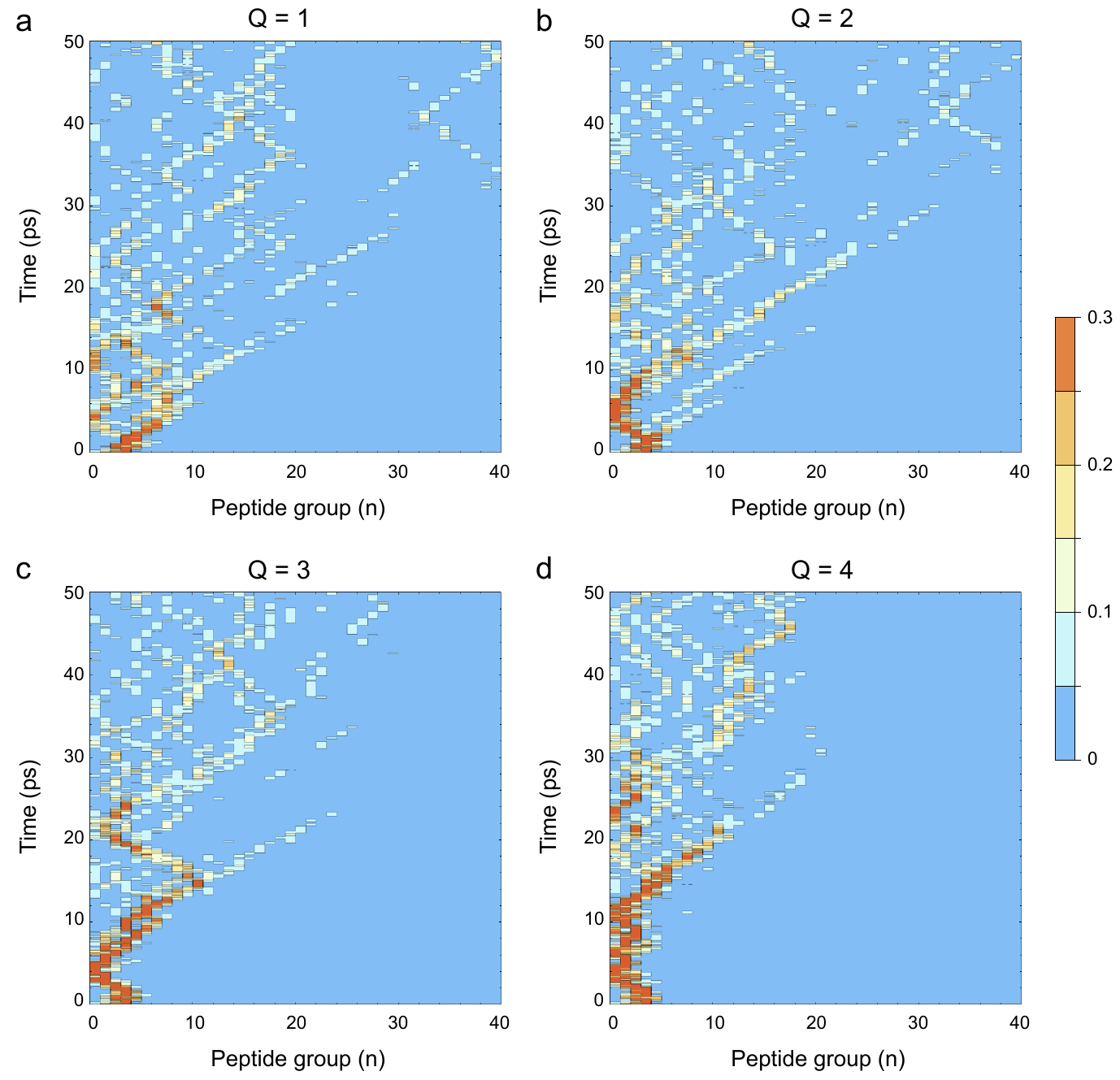}
\par\end{centering}

\caption{\label{fig:4}Dynamics of a soliton inside a single $\alpha$-helix
spine with $\tilde{M}=1M$ and $\tilde{w}=1w$ at $T=310$~K, with
different number $Q$ of amide~I energy quanta visualized through
the exciton quantum probability $|a_{n}|^{2}$ at each peptide group~$n$.
The~Gaussian pulse of amide~I energy is applied at the N-end of a protein
$\alpha$-helix spine with $n_{\max}=40$ peptide groups. The soliton
is pinned and depending on the value of $Q$ persists for 4--20~ps before being dispersed by the thermal noise.}
\end{figure*}

In the model of a single $\alpha$-helix spine with $\tilde{M}=1M$
and $\tilde{w}=1w$, the effects of the thermal noise were devastating
to soliton stability and its biological utility (Fig.~\ref{fig:4}).
In order to assess the effect of increasing the number $Q$ of amide~I quanta on soliton stability, we have used the same set $S_{0}$
of Wiener processes and the same initial thermalized state $|\psi_{\textrm{ph}}(0)\rangle$
of the protein lattice. For $Q=1$, the soliton persisted for only
$\approx4$~ps (Fig.~\ref{fig:4}a), for $Q=2$ it persisted a bit
longer for $\approx10$~ps (Fig.~\ref{fig:4}b), and for $Q\geq3$
the soliton lifetime was $\approx20$~ps (Figs.~\ref{fig:4}c,d).
The most important problem in the simulations with $Q\geq2$, however,
was that the soliton remained pinned at the site of its creation and
did not move along the $\alpha$-helix spine (Fig.~\ref{fig:4}b-d).
In other words, the thermal noise was unable to set the soliton in
motion, not even in the fashion of a random walk where the soliton
is wandering aimlessly around. Although the pinning for $Q\geq3$
is already predictable by the simulations at zero temperature (Figs.~\ref{fig:2}c,d),
comparison of the case with $Q=2$ at zero and at
physiological temperature (Fig.~\ref{fig:2}b vs Fig.~\ref{fig:4}b)
shows that the presence of thermal noise is actually capable of pinning
the soliton that otherwise would be in motion.
These findings for a single isolated $\alpha$-helix spine are qualitatively
different from the simulations with the full protein $\alpha$-helix
(elaborated in the next subsection) and are remarkable
in Davydov's theory of protein solitons for two reasons:
First, they reproduce the ultrashort soliton lifetimes for $Q=1$ in single spine found by Lomdahl and Kerr \cite{Lomdahl1985}.
Second, they reproduce the soliton ineffectiveness for transportation of energy in single spine for $Q\geq2$ \cite{Lomdahl1990}.
Thus, we are in complete agreement with previous research done on
a single isolated $\alpha$-helix spine. But the main conclusion
to be drawn is that the model of a single $\alpha$-helix spine is
not a trustworthy model for studying the transport of energy in intact protein $\alpha$-helices.
In particular, it is inadequate in handling the stochastic quantum
dynamics of proteins introduced by the presence of physiological thermal noise.

\subsubsection{Massive single-chain model of protein \texorpdfstring{$\alpha$}{alpha}-helix}

\begin{figure*}[t]
\begin{centering}
\includegraphics[width=140mm]{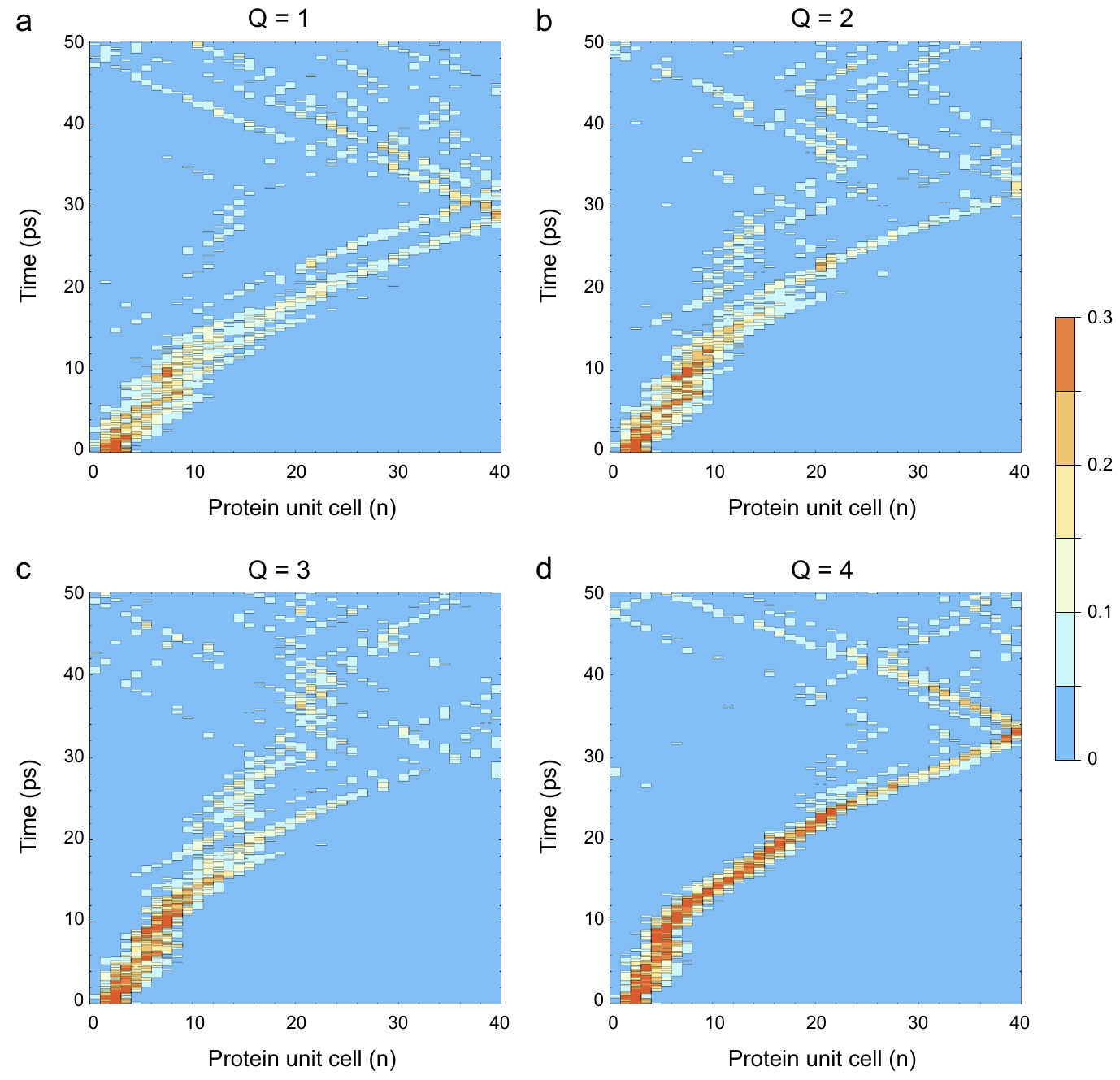}
\par\end{centering}

\caption{\label{fig:5}Dynamics of a soliton moving inside the massive single-chain model of a protein
$\alpha$-helix with $\tilde{M}=3M$ and $\tilde{w}=3w$ at $T=310$~K, with different number $Q$ of amide~I energy quanta visualized
through the exciton quantum probability $|a_{n}|^{2}$ at each protein
unit cell~$n$. The Gaussian pulse of amide~I energy is applied at
the N-end of a protein $\alpha$-helix with $n_{\max}=40$ unit cells,
each of which is comprised of three peptide groups. The soliton
persists for 30--40~ps in the presence of thermal noise.}
\end{figure*}

In the massive single-chain model of a protein $\alpha$-helix with $\tilde{M}=3M$
and $\tilde{w}=3w$, the presence of thermal noise (based on the set $S_0$ of Wiener processes) at physiological
temperature, $T=310$~K, also introduced dynamic instability but its
action was delayed, namely, it disintegrated the soliton with lifetime
of over $30$~ps (Fig.~\ref{fig:5}). For $Q\leq3$, the thermal noise
smeared the soliton trajectory in the initial $20$~ps of the simulation,
after which the soliton bifurcated into two tracks with smaller amplitudes.
Eventually, the soliton was dispersed after $\approx30$~ps (\ref{fig:5}a-c).
For $Q=4$, the self-trapping effect was strong enough to allow the
soliton to travel from the N-end along the whole extent of the protein
$\alpha$-helix and reach the opposite C-end. In addition to the enhanced
stability, the lifetime of the soliton was also extended to $\approx40$
ps (\ref{fig:5}d). Thus, the larger mass~$\tilde{M}$ of the full
protein $\alpha$-helix and the higher number~$Q$ of amide~I exciton
quanta cooperate together to enhance the soliton stability against
thermal noise.

\begin{figure*}[t]
\begin{centering}
\includegraphics[width=140mm]{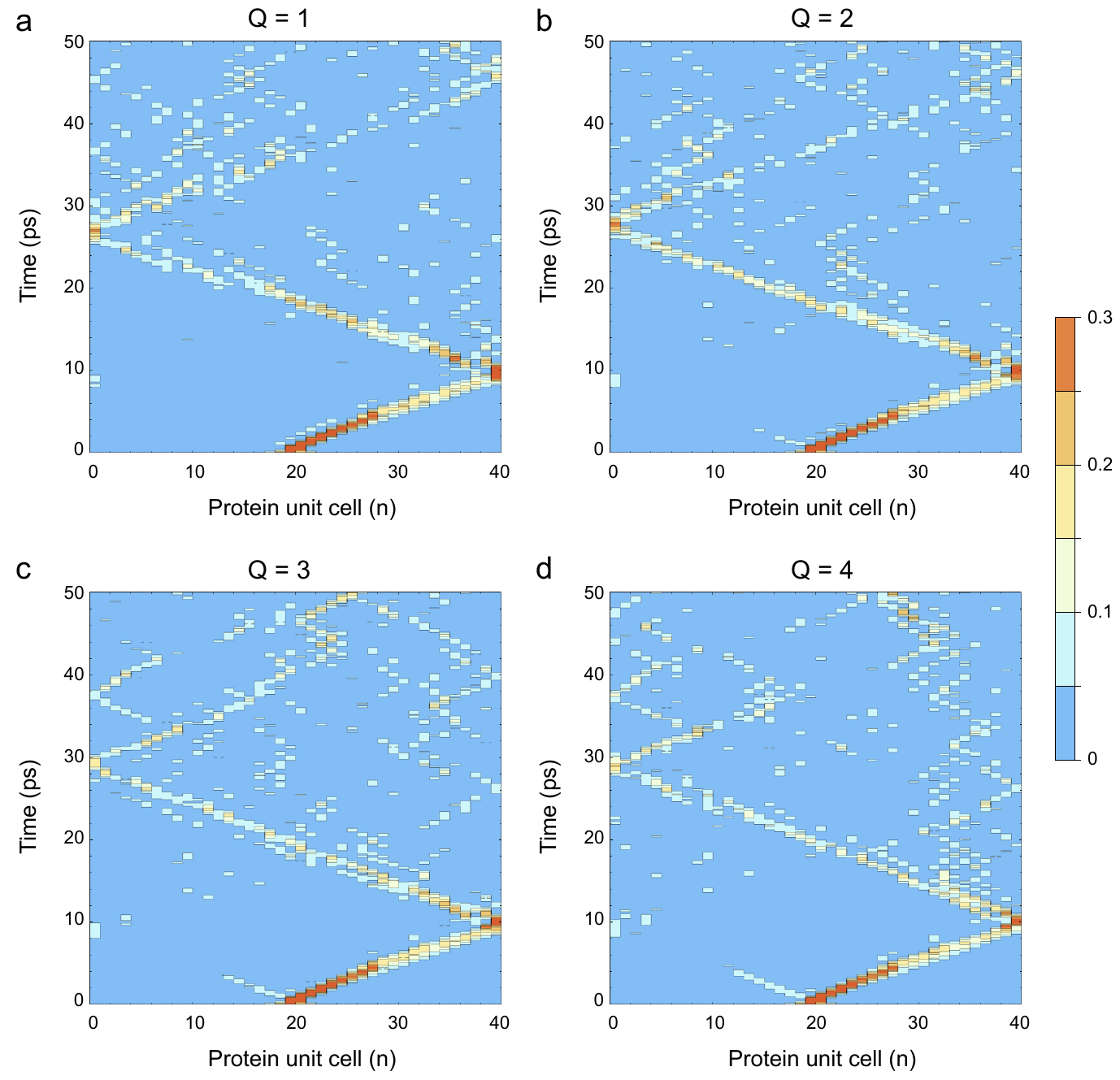}
\par\end{centering}

\caption{\label{fig:6}Dynamics of a soliton moving inside the massive single-chain model of a protein
$\alpha$-helix with $\tilde{M}=3M$ and $\tilde{w}=3w$ at $T=310$~K, with different number~$Q$ of amide~I energy quanta visualized
through the exciton quantum probability $|a_{n}|^{2}$ at each protein
unit cell~$n$. The Gaussian pulse of amide~I energy is applied in
the middle of a protein $\alpha$-helix with $n_{\max}=40$ unit cells,
each of which is comprised of three peptide groups. The soliton
persists for 20--30~ps in the presence of thermal noise.}
\end{figure*}

\pagebreak
Gaussian pulses of amide~I energy were also capable to generate moving
solitons when delivered in the middle of the protein $\alpha$-helix
(Fig.~\ref{fig:6}). Because in this instance the soliton was able to
reach the end of the helix within 10~ps, the simulations showed soliton
bouncing from the C-end and propagation in reversed direction towards
the N-end. For $Q\leq3$, the soliton survived for 30~ps and reached
the N-end of the helix (Fig.~\ref{fig:6}a-c). For $Q=4$, however,
the bounce from the C-end led to splitting of the soliton into two
tracks, only one of which reached the N-end (Fig.~\ref{fig:6}d).
Because the soliton consists of the expectation values of detecting
the amide~I excitons at different lattice sites, splitting of the
soliton indicates splitting of the quantum probability for utilization
of amide~I quanta. In other words, in the presence of multiple soliton
tracks, the delivery of metabolic energy to different sites in the
protein $\alpha$-helix becomes a probabilistic event.

\begin{figure*}[t!]
\begin{centering}
\includegraphics[width=140mm]{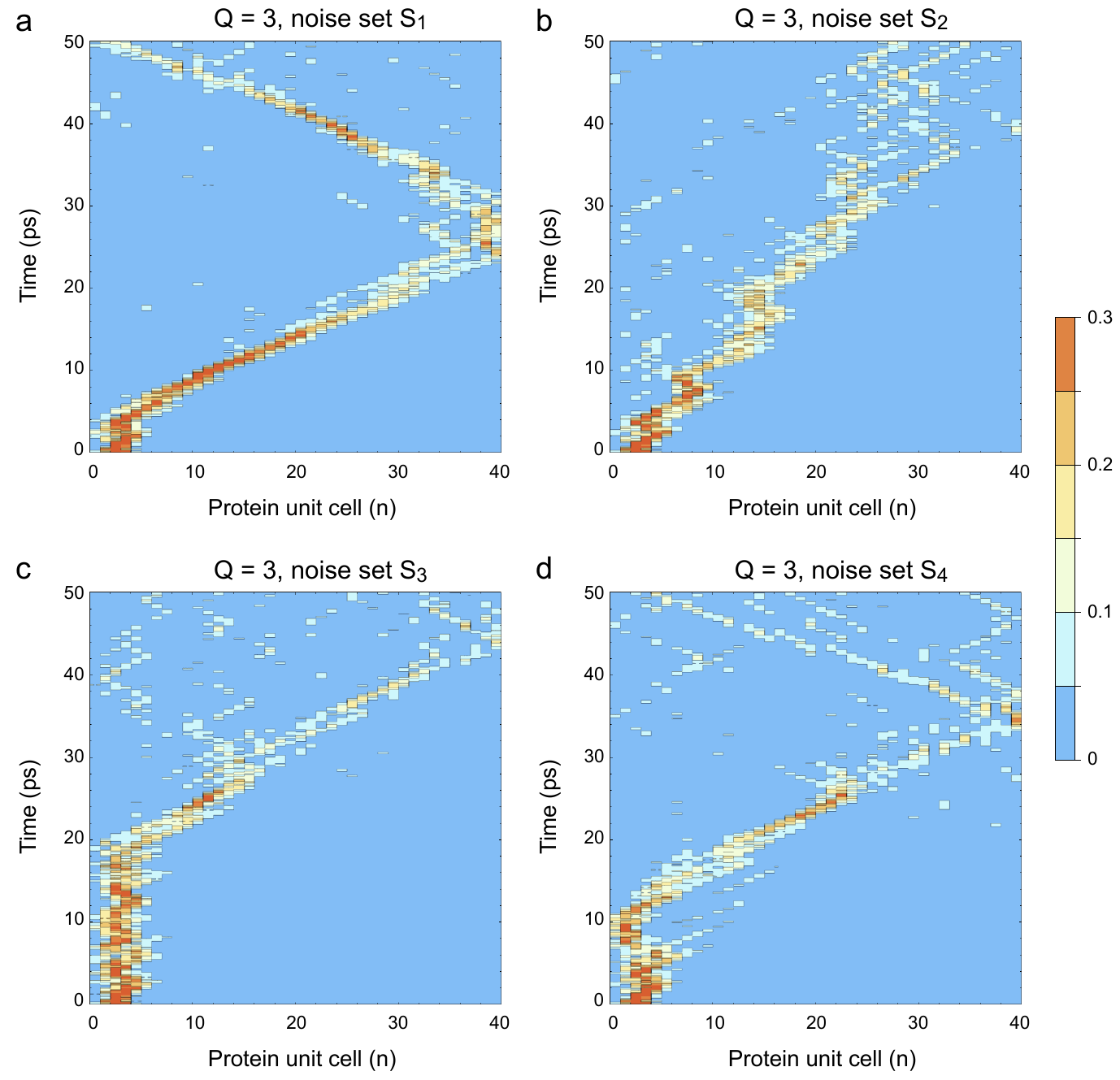}
\par\end{centering}

\caption{\label{fig:7}Dynamics of a soliton moving inside the massive single-chain model of a protein
$\alpha$-helix with $\tilde{M}=3M$ and $\tilde{w}=3w$ at $T=310$~K,
for $Q=3$ amide~I energy quanta under the influence of different
sets $S_{i}$ of Wiener processes, visualized through the exciton
quantum probability $|a_{n}|^{2}$ at each protein unit cell $n$.
The Gaussian pulse of amide~I energy is applied at the N-end of a
protein $\alpha$-helix with $n_{\max}=40$ unit cells, each of which
is comprised of three peptide groups. The protein soliton starts from
the same initial thermalized state of the lattice but exhibits different
trajectories for different actualizations $S_{i}$ of the thermal
noise.}
\end{figure*}

To further obtain unbiased statistical data on the stochastic quantum
dynamics of solitons in the full protein $\alpha$-helix with $\tilde{M}=3M$
and $\tilde{w}=3w$ at $T=310$~K, we have randomly selected another
different initial thermalized state $|\psi_{\textrm{ph}}(0)\rangle$ of the lattice
and generated 20~different sets $S_{i}$ of Wiener processes, $i\in\{1,2,\ldots,20\}$.
Then, we have computed soliton plots with $Q=3$ amide~I exciton quanta,
which corresponds to the energy released by a single ATP molecule
in physiological conditions \cite{Barclay2020,Kammermeier1982,Weiss2005}.
As expected by the stochastic nature of the Langevin dynamics introduced
by the thermal noise, the protein soliton followed different paths
for different simulation runs (Fig.~\ref{fig:7}). In particular,
the traversing of the entire extent of the protein $\alpha$-helix was a
probabilistic event that occurred with different initial delay time,
which was $\approx5$~ps in the noise sets $S_{1}$ and $S_{2}$ (Figs.
\ref{fig:7}a,b), $\approx10$~ps in the noise set $S_{4}$ (Fig.
\ref{fig:7}d), and $\approx20$~ps in the noise set $S_{3}$ (Fig.
\ref{fig:7}c). Overall, the observed mean$\pm$standard deviation
of the soliton lifetime was $33.2\pm8.8$~ps and the average soliton
speed until the time point of soliton dispersal was $394\pm198$ m/s.
Compared with the speed of $439$ m/s recorded at zero temperature
(Fig.~\ref{fig:3}c), the presence of thermal noise decreased by $\approx10\%$
the average soliton speed due to existing brief periods of time during
which the soliton appeared to be temporarily pinned. Nevertheless, there were
also periods of time during which the soliton speed was boosted due to ambient thermal noise, as evidenced by the large standard deviation that is $\approx50\%$ of the measured average soliton speed.

Taken together, the above results demonstrate that the cooperative excitation
of the three $\alpha$-helix spines with $Q=3$ amide~I energy quanta
is able to counteract the negative effects of the thermal noise for $\approx30$~ps
during which time the soliton could successfully traverse the whole
extent of an 18-nm-long protein $\alpha$-helix. Since the energy released by a single ATP molecule
suffices to initiate $Q=3$ amide~I exciton quanta, it is reasonable to expect that the protein
solitons are sustained by the available metabolic energy despite the presence of
ambient thermal noise. The feasibility of protein solitons as means for transportation
of energy in biological systems is further supported by the general agreement
between the simulated soliton lifetime of about $30$~ps and the
experimentally measured lifetime of amide~I vibrational modes in sperm
whale myoglobin, which has been reported to be at least $15$~ps \cite{Xie2000}.
It should be noted, however, that
the average length of the eight \mbox{$\alpha$-helices} of myoglobin is only 2.2~nm and the longest $\alpha$-helix is 3.8~nm \cite{Phillips1980}. Thus, the solitons can reach 4--5 times faster the opposite $\alpha$-helix end where they can be absorbed or suffer amplitude loss during bouncing. Consequently, the lifetime of the soliton as measured in globular proteins might not be indicative of the lifetime in any longer $\alpha$-helices, such as those found in transmembrane proteins or in ion channels \cite{Georgiev2020c,Kariev2021}.

\section{Three-spine model with lateral coupling}
\label{sec:5}

Having demonstrated that the massive single-chain model of a protein $\alpha$-helix exhibits greatly enhanced thermal stability,
we have further investigated the more elaborate three-spine Davydov model \cite{Davydov1982}, which has been subject to intense
scrutiny in 1990s. Through careful examination of the original articles, we discovered that the highly
influential works by A. C. Scott \cite{Scott1982,Scott1992} contain
a biologically inadequate Hamiltonian for the phonon lattice in which only
the longitudinal coupling by hydrogen bonds is considered, but there
is no lateral coupling between the peptide groups, which are provided
by the covalent C-C and C-N bonds by the protein backbone spiral.
This omission has far-reaching biological consequences because Scott's model effectively
describes three uncoupled spines much like our isolated $1M$ spine model
and not like the massive single-chain model of a protein $\alpha$-helix.
We point out that Davydov's original Hamiltonian \cite[Eq.~(3.3)]{Davydov1982}
actually contains lateral couplings between the lattice sites and
these are considered to be stronger than the longitudinal couplings
by hydrogen bonds. The general three-spine Hamiltonian is the sum
of three parts
\begin{eqnarray}
\hat{H}_{\textrm{ex}} &=& \sum_{n,\alpha}\left[E_{0}\hat{a}_{n,\alpha}^{\dagger}\hat{a}_{n,\alpha}-J_{1}\left(\hat{a}_{n,\alpha}^{\dagger}\hat{a}_{n+1,\alpha}+\hat{a}_{n,\alpha}^{\dagger}\hat{a}_{n-1,\alpha}\right)+J_{2}\left(\hat{a}_{n,\alpha}^{\dagger}\hat{a}_{n,\alpha+1}+\hat{a}_{n,\alpha}^{\dagger}\hat{a}_{n,\alpha-1}\right)\right] \\
\hat{H}_{\textrm{ph}} &=& \frac{1}{2}\sum_{n,\alpha}\left[\frac{\hat{p}_{n,\alpha}^{2}}{M}+w_{1}\left(\hat{u}_{n+1,\alpha}-\hat{u}_{n,\alpha}\right)^{2}+w_{2}\left(\hat{u}_{n,\alpha+1}-\hat{u}_{n,\alpha}\right)^{2}\right] \label{eq:lattice-3}\\
\hat{H}_{\textrm{int}} &=& \sum_{n,\alpha}\left[\chi_{r}\hat{u}_{n+1,\alpha}+\left(\chi_{l}-\chi_{r}\right)\hat{u}_{n,\alpha}-\chi_{l}\hat{u}_{n-1,\alpha}\right]\hat{a}_{n,\alpha}^{\dagger}\hat{a}_{n,\alpha}
\end{eqnarray}
where $J_{1}=967.4$~$\mu$eV is the nearest neighbor dipole-dipole
coupling energy along a spine, $J_{2}=1535.4$~$\mu$eV is the
nearest lateral neighbor dipole-dipole coupling energy between spines,
$w_{1}=13$ N/m is the spring constant of the longitudinal hydrogen
bonds in the lattice of peptide groups, $w_{2}$ is the spring constant
of the lateral coupling of the lattice of peptide groups due to covalent
bonds in the protein backbone, and $\alpha\in\left\{ 0,1,2\right\} $
is an index denoting each of the three spines in the protein $\alpha$-helix.
To keep our notation concise in the Hamiltonians and avoid writing
$\textrm{mod}3$ in subscripts, we consider it implicitly understood
that the index $\alpha$ is subject to modulo~3 arithmetic, namely
$\alpha+1=3\equiv0$ and $\alpha-1=-1\equiv2$, with $\equiv$ denoting
the congruence modulo~3 relation.

\begin{figure*}[t!]
\begin{centering}
\includegraphics[width=140mm]{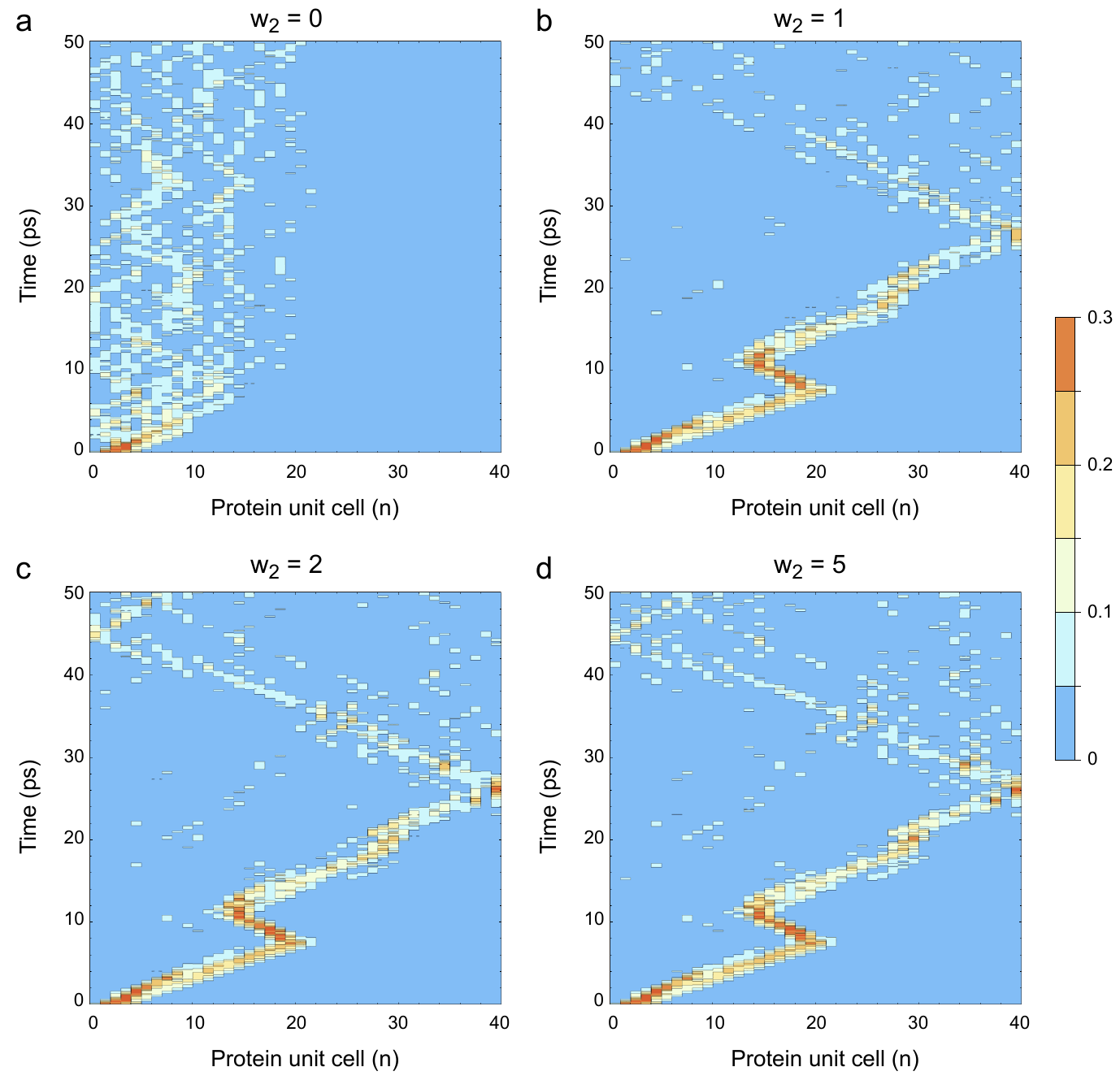}
\par\end{centering}

\caption{\label{fig:8}Dynamics of a soliton moving inside the three spine model of
a protein $\alpha$-helix with different strength of the lateral coupling $w_2$ (measured in units of $w_1$) at $T=310$~K,
for $Q=3$ amide~I energy quanta under the influence of a fixed set of 120 Wiener processes, visualized through the exciton
quantum probability $\sum_\alpha |a_{n,\alpha}|^{2}$ at each protein unit cell $n$.
The Gaussian pulse of amide~I energy is applied at the N-end of a
protein $\alpha$-helix with $n_{\max}=40$ unit cells, each of which
is comprised of three peptide groups. The protein soliton starts from
the same initial thermalized state of the lattice.}
\end{figure*}

By setting $w_{2}=0$, we obtain exactly Scott's model \cite[Eq.~(1.3)]{Scott1992}
in which the three spines are uncoupled. In contrast, by choosing
different values $w_{2}\geq w_{1}$, we obtain Davydov's original
three spine model with coupled spines. Using the ansatz \eqref{eq:ansatz} and the generalized
Ehrenfest theorem, the system of stochastic Langevin equations of
motion for the full three-spine model becomes
\begin{align}
\imath\hbar\frac{d}{dt}a_{n,\alpha} & =  -J_{1}\left(a_{n+1,\alpha}+a_{n-1,\alpha}\right)+J_{2}\left(a_{n,\alpha+1}+a_{n,\alpha-1}\right)+\left[\chi_{r}b_{n+1,\alpha}+(\chi_{l}-\chi_{r})b_{n,\alpha}-\chi_{l}b_{n-1,\alpha}\right]a_{n,\alpha}\nonumber\\ 
\label{eq:full-1} \\
\frac{d^{2}}{dt^{2}}b_{n,\alpha} &= \frac{w_1}{M}\left(b_{n-1,\alpha}-2b_{n,\alpha}+b_{n+1,\alpha}\right)+\frac{w_2}{M} (b_{n,\alpha-1}-2b_{n,\alpha}+b_{n,\alpha+1})\nonumber\\
&\quad -\frac{Q}{M}\left[\chi_{r}\left|a_{n-1,\alpha}\right|^{2}+(\chi_{l}-\chi_{r})\left|a_{n,\alpha}\right|^{2}-\chi_{l}\left|a_{n+1,\alpha}\right|^{2}\right]
 -\Gamma\frac{d}{dt}b_{n,\alpha}+\sqrt{\frac{2\Gamma k_{B} T}{M}}\frac{d}{dt}W_{n,\alpha}(t)
\label{eq:full-2}
\end{align}

Computer simulations were then performed with an initial symmetric discrete Gaussian pulse of amide~I energy, which was equally distributed over all three spines:
\begin{equation}
\frac{1}{\sqrt{3}} e^{\imath\alpha\Delta\omega_2}
\left\{ A_{2,\alpha}e^{-\imath2\Delta\omega_{1}},A_{1,\alpha}e^{-\imath\Delta\omega_{1}},A_{0,\alpha},A_{1,\alpha}e^{+\imath\Delta\omega_{1}},A_{2,\alpha}e^{+\imath2\Delta\omega_{1}}\right\} 
\label{eq:pulse-3}
\end{equation}
where $A_{0,\alpha}=\sqrt{0.322}$, $A_{1,\alpha}=\sqrt{0.24}$, $A_{2,\alpha}=\sqrt{0.099}$,
and the phase modulation was present along the spines $\Delta\omega_1=\frac{\pi}{12}$, but not laterally between the spines $\Delta\omega_2=0$.
The solitons were visualized by the quantum probability of
finding the amide~I exciton inside the $n$th protein unit cell
\begin{equation}
P\left(a_{n}\right)=\sum_{\alpha}|a_{n,\alpha}|^{2}=|a_{n,0}|^{2}+|a_{n,1}|^{2}+|a_{n,2}|^{2}
\end{equation}

At zero temperature, $T=0$~K, the soliton plots for $Q=3$ are indistinguishable
from the simulation shown in Fig.~\ref{fig:3}c, with soliton speed of 439~m/s.
The numerical differences between the massive single-chain model of a protein $\alpha$-helix and the three-spine model are negligible
as they are on the order of $5\times10^{-7}$. Due to the symmetric
initial conditions, the soliton plots of the three-spine
model at zero temperature are identical for all values of the lateral coupling spring
constant~$w_{2}$ (see \ref{app-F}).

The lateral coupling $w_{2}$ in the protein $\alpha$-helix, however, is of
paramount importance in the presence of thermal noise at physiological
temperature, $T=310$~K. In order to compare the effects of varying
$w_{2}$, we have generated a set of $3\times40=120$ Wiener processes
and pre-thermalized the protein lattice in the absence of amide~I
excitons, $Q=0$. Then, we have performed computer simulations with
exactly the same initial thermalized state of the lattice and the
same set of 120 Wiener processes.

In the absence of lateral coupling, $w_{2}=0$,
the model has the properties of three uncoupled spines
of hydrogen bonds and the soliton is dispersed rapidly within 2~ps
(Fig.~\ref{fig:8}a).
This is consistent with the results of poor thermal stability
of solitons in the single isolated spine (Fig.~\ref{fig:4}a) and reproduces the findings by Lomdahl and Kerr \cite{Lomdahl1990}.
Therefore, Scott's model without lateral coupling is simply a triplet of single isolated spines, each of which lacks thermal stability, and hence is not a model of intact protein $\alpha$-helix.
When the lateral coupling $w_{2}$ is introduced, at values of $1\times$, $2\times$
or $5\times$ the strength of the hydrogen bonds $w_{1}$, the thermal
stability of the protein soliton is drastically increased to 30~ps
(Fig.~\ref{fig:8}b-d). This result is consistent with the findings obtained
from the massive single-chain model of a protein $\alpha$-helix
(Fig.~\ref{fig:7}).

The observed thermal stability of the three-spine model with lateral coupling is corroborated by recent computational study by Brizhik and collaborators \cite{Brizhik2019}, where an elaborate spiral protein lattice with greater number of lateral couplings is considered. In the latter model, each peptide group is coupled to 6 neighboring peptide groups in the spiral, whereas in our three-spine Hamiltonian \eqref{eq:lattice-3}, each peptide group is coupled to only 4 neighboring peptide groups. Because we were still able to obtain extended soliton lifetime, we consider our three-spine Hamiltonian as the minimal model that guarantees thermal stability.
Further inclusion of additional (diagonal) lateral couplings can only reinforce the model against thermal noise \cite{Brizhik2019}.

Previous work by Cruzeiro-Hansson and Takeno classified the Davydov model in three regimes: the quantum
regime, in which both the exciton and the lattice are treated quantum mechanically; the mixed quantum-classical
regime, in which the exciton is treated quantum mechanically but the lattice is considered classical;
and the classical regime, in which both the exciton and the lattice are treated classically \cite{Cruzeiro1996a,Cruzeiro1996b,Cruzeiro1997}.
Within this classification, our treatment is fully quantum because we use only the Schr\"{o}dinger equation \eqref{eq:schrodinger} to derive the equations of motion for both the excitons and the expectation values of the lattice displacements.
Our derivations, outlined in detail in \cite{Georgiev2020a}, are formally equivalent to the full quantum mechanical derivations by Kerr and Lomdahl \cite{Kerr1987,Kerr1990}.

The fundamental difference between the classical and the quantum treatment of the lattice is not the mathematical form of \eqref{eq:full-2}, but its physical interpretation: In classical physics based on Hamilton's equations, the physical quantities $b_{n,\alpha}$ represent the \emph{lattice displacements} from their equilibrium positions.
This means that the classical Davydov soliton with given exciton distribution has a fixed lattice deformation that is responsible for the self-trapping effect. In quantum physics based on the Schr\"{o}dinger equation, however, the physical quantities $b_{n,\alpha}$ represent the \emph{expectation values} of the lattice displacements from their equilibrium positions. Only when we average over all measurement outcomes for the lattice displacements, we will obtain $b_{n,\alpha}=\langle \Psi(t)| \hat{u}_{n,\alpha} | \Psi(t) \rangle$.
This means that the quantum Davydov soliton with given exciton distribution does not have a fixed lattice displacement, but is in so-called Glauber state, which is a quantum coherent superposition of different lattice deformations \eqref{eq:coherent} \cite{Glauber1963a,Glauber1963b}. Therefore, the quantum nature of the Davydov soliton is manifested in the fact that the exciton dynamics \eqref{eq:full-1} does not depend on the possible actualization of a \emph{particular lattice displacement} upon quantum measurement, but depends on the \emph{expectation value} from all possible lattice displacements that could be measured.

Another point where our work differs from the one by Cruzeiro-Hansson and Takeno \cite{Cruzeiro1996a,Cruzeiro1996b,Cruzeiro1997} is their claim that the introduction of a Langevin thermal bath coupled to the protein lattice necessarily forces the whole system to behave classically. Their observation of rapid dispersal of solitons in a single isolated spine cannot be used for justification of the latter claim because the use of lattice Hamiltonian lacking lateral coupling does not provide a valid model of intact protein $\alpha$-helix. Furthermore, our demonstration of extended soliton lifetime of 30~ps in the presence of a Langevin thermal bath shown in Fig.~\ref{fig:8} clearly contradicts the hypothesis that the mere presence of Langevin terms is detrimental for Davydov's model.
A~mathematically rigorous derivation of Langevin molecular dynamics, for the case of a composite quantum system that is weakly coupled to a heat bath of many fast particles, has been recently developed by Hoel and Szepessy \cite{Hoel2020}. Thus, we maintain the view that the system of equations \eqref{eq:full-1} and \eqref{eq:full-2} represents a quantum Langevin model \cite{Ford1987,Araujo2019,deOliveira2020} of protein $\alpha$-helix immersed in water-based solvent at physiological temperature.

It is interesting to note that Savin and Zolotaryuk in \cite{Savin1993} have previously
modified Davydov's model of single spine by introducing 
an on-site potential that accounts for the effects of the other two
lateral spines, thereby enhancing the soliton's thermal stability.
This latter model is also simplified similarly to our massive single-chain model of a protein $\alpha$-helix, which shows that the stabilizing effect
of the extra on-site potential could be mimicked by increased (tripled)
mass of the individual lattice sites. Furthermore, because the added
on-site potential in the Savin--Zolotaryuk model has a spring constant
$K=30.5$ N/m, it effectively describes the behavior of a protein
unit cell from the three-spine model with lateral coupling
$w_{2}\approx 2w_{1}$ for symmetric initial distribution of the Gaussian
amide~I energy pulse. This latter limitation, however, is also shared
by the massive single-chain model of a protein $\alpha$-helix, which
describes only symmetric solitons. Instead, for studying the quantum
dynamics of asymmetric initial distributions of the Gaussian amide~I energy, one needs to use the full three-spine model with lateral
coupling and solve the three times larger system of stochastic differential
equations \eqref{eq:full-1} and \eqref{eq:full-2}.

Taken together, the results obtained from both the massive single-chain model of a protein $\alpha$-helix and the three-spine model support
the thermal stability of protein solitons for a biologically feasible
period of time exceeding 30~ps. Furthermore, they pinpoint the omission
of lateral couplings in the three-spine lattice as a potent factor
that disintegrates the soliton in the presence of thermal noise thereby
undermining the veracity of any inferences with regard to biological
utility of the protein solitons.

\section{Concluding remarks}

Protein functions involve physical work and as such can be performed
only at the expense of free energy released by biochemical reactions.
The hydrolysis of a single ATP molecule is sufficient to initiate
three amide~I quanta, which could then propagate along the
protein \mbox{$\alpha$-helices} and be utilized for the execution
of various physiological activities. The quantum system of equations of motion
\eqref{eq:gauge-1} and \eqref{eq:gauge-2} for amide~I energy quanta
coupled to lattice phonons in proteins could be analyzed within the continuum approximation in order to derive self-trapped
propagating quasiparticles referred to as solitons \cite{Davydov1976,Davydov1979,Brizhik1983,Brizhik1988,Malomed2020,Vakhnenko2021,Luo2021}.
The lifetime of these solitons is potentially infinite at zero temperature,
$T=0$~K, which allows for analytic derivation of soliton energy,
effective mass, and speed of energy transportation \cite{Davydov1981,Kivshar1989,Brizhik1993,Brizhik2004,Georgiev2020b}.
Realistic biological systems, however, operate at physiological temperature,
$T=310$~K, at which the ambient thermal noise exerts a disrupting effect
on the soliton's stability \cite{Lomdahl1985,Lomdahl1990,Cruzeiro1988,Cruzeiro1990,Forner1990,Forner1991,Forner1991b,Forner1991c}.
Therefore, in this study we have explored the thermal stability
of protein solitons in a biological context using a Langevin modification
\eqref{eq:SDE} of the quantum system of equations of motion.

Computer simulations, based on a set of biophysical parameter values
established for proteins, have demonstrated that the soliton lifetime
at physiological temperature is strongly dependent on the number~$Q$
of initiated amide~I exciton quanta and the type of theoretical model
chosen, namely, whether a single $\alpha$-helix spine is investigated
in isolation or the full protein $\alpha$-helix with three coupled \mbox{$\alpha$-helix}
spines is considered. In the model of a single isolated $\alpha$-helix
spine, the solitons disintegrate quickly by the ambient thermal noise
with a lifetime of several picoseconds, which does not suffice for
transportation of energy at a biologically meaningful distance. In
the model of the full protein $\alpha$-helix, however, the presence
of $Q=3$ amide~I quanta ensures a cooperative stabilizing effect
upon the formed soliton for a time period over $30$~ps, which
is long enough for the soliton to traverse the full extent of an
$\alpha$-helix with $n_{\max}=40$ lattice sites.
Thus, our computational results establish that multiquantal solitons
in full protein $\alpha$-helices with three $\alpha$-helix spines
possess the required thermal stability in order to support energy transportation in proteins of living systems \cite{Brizhik2006,Brizhik2019}.
This robustness against thermal noise of the generalized Davydov model with lateral coupling
also lends indirect support for the predicted importance of topological solitons for protein folding in relation to their biological
function \cite{Krokhotin2012,Peng2020}.

Here, our main goal was to characterize the conditions under which thermal stability of Davydov solitons is achieved.
The presented three-spine Davydov model indeed highlights the minimal engineering of the quantum Hamiltonian that supports thermally stable solitons. By adding extra coupling parameters and considering vibrational motion in all 3~spatial dimensions, however, one might be able to achieve even better performance. The possibility of further tweaking and upgrading the model quantum Hamiltonian, we leave for future investigations.

\appendix

\section{Soliton width}
\label{app-A}

\setcounter{equation}{30}

\counterwithout{figure}{section}
\counterwithout{equation}{section}

The biological feasibility of Davydov's model has been heavily criticized
in 1990s by different authors using a combination of suboptimal parameter
choices including extreme narrowing of the initial soliton width to
a single lattice site, namely, triggering the soliton with a bare exciton.
In previous exploratory works, we have shown that wider solitons exhibit higher stability
and move at lower speeds \cite{Georgiev2019a,Georgiev2020a}. This suggests that
biological systems may harness soliton stability by applying the amide~I energy
through wide Gaussian pulses extending over three to five lattice
sites (peptide groups) along each $\alpha$-helix spine. In fact,
biochemical data unambiguously shows that the free energy from ATP
hydrolysis is not released by a single event of pyrophosphate bond
cleavage, but released sequentially by several molecular events involving
destruction and formation of ionic or hydrogen bonds in the hydration shell of the Mg\textsuperscript{2+} ion.
The free energy value of 0.63 eV used throughout this work is for
the hydrolysis of Mg\textsuperscript{2+}-ATP to Mg\textsuperscript{2+}-ADP
\cite{Wu2008} as described by the following chemical reaction:

\begin{center}
\scalebox{0.65}{
\schemestart[]
\raisebox{4em}{\chemfig{OH-[:90,,1]-[:58.5,1.182](-[:92.7,1.395]N-[:165]=^[:93](-[:153](-[:93,,,1]NH_2)=^[:213]N-[:273]=^[:333]N-[:33])-[:21]N=_[:309]-[:237]\phantom{N})-[:151.8,1.176]O-[:207.2,1.11](-[:292.1,1.082](-[:265,,,1]OH)-[:357])-[:108.2,0.485]-[:182.8,0.779]O-[:180]P(=[:90]O)(-[:270]\mcfright{O_\alpha}{^{\mcfminus}})-[:180]O-[:180]P(=[:90]O)-[:180]O-[:180]P(=[:90]O)(-[:180,,,2]^{\mcfminus}O)-[:270]O_\gamma^{\mcfminus}-[:316.2,1.491]Mg^{2+}-[:48.2,1.386]O_\beta^{\mcfminus}(-[:90]\phantom{P})}}
\arrow{0}[,0] \+ \arrow{0}[,0]
\chemfig{H_2O}
\arrow{->} \arrow{0}[,0]
\chemfig{OH-[:180,,1]P(=[:90]O)(-[:270]\mcfright{O_\gamma}{^{\mcfminus}})-[:180,,,2]HO}
\arrow{0}[,0] \+ \arrow{0}[,0]
\raisebox{4em}{\chemfig{OH-[:90,,1]-[:58.5,1.182](-[:87.5,1.572]N-[:165]=^[:93](-[:153](-[:93,,,1]NH_2)=^[:213]N-[:273]=^[:333]N-[:33])-[:21]N=_[:309]-[:237]\phantom{N})-[:151.8,1.176]O-[:207.2,1.11](-[:292.1,1.082](-[:265,,,1]OH)-[:357])-[:106.7,0.528]-[:183.2,0.69]O-[:180]P(=[:90]O)-[:180]O-[:180]P(=[:90]O)(-[:180,,,2]^{\mcfminus}O)-[:270]O_\beta^{\mcfminus}-[:317.4,1.33]Mg^{2+}-[:41.4,1.36]O_\alpha^{\mcfminus}(-[:90]\phantom{P})}}
\schemestop
}
\end{center}

After the hydrolysis of the outer pyrophosphate bond, the Mg\textsuperscript{2+}
ion detaches from the $\gamma$-phosphate and attaches to the $\alpha$-phosphate,
while retaining its ionic bond to the $\beta$-phosphate
\cite{Barrozo2017}. The linear extension of the four negatively charged
O$^{-}$ atoms in the three phosphate groups exceeds 1~nm after the cleavage,
which is consistent with a Gaussian energy pulse delivered onto
three to five C=O groups in the protein $\alpha$-helix. In fact,
the Gaussian pulse that we have used in the present simulations contains
\textgreater{}80\% of the amide~I energy concentrated over the middle
three C=O groups from the total of five.

The wavelength of a single infrared photon with energy 0.63~eV is
1968 nm. Because the span of five peptide groups along the spine is
only 2.25 nm, there has to be a mechanism that focuses the pulse of
free energy released from ATP onto the receptive protein $\alpha$-helix.
Furthermore, the efficiency of this physical process should be high, thus
allowing the protein to utilize locally the ATP energy for accomplishing
biological functions with almost unit probability.
Due to the large wavelength and lack of infrared lensing mechanism in living systems,
it is unlikely that the ATP energy is absorbed by the protein in the
form of infrared photons. Instead, it is more plausible that the physical
delivery of free energy results from direct interaction, e.g. of the four electrically
charged O$^{-}$ atoms from the ATP phosphate groups and several (three to five) C=O groups
in the protein $\alpha$-helix.

Alexander Davydov himself repeatedly emphasized the importance of the soliton
width being extended over several C=O groups for the enhancement of
soliton stability \cite{Davydov1986,Davydov1991}. In fact, the localization
of the amide~I energy onto a single site removes the phonon dressing
and produces a bare exciton that moves with high speed and easily
loses amplitude even at zero temperature \cite{Georgiev2019a} (see also \ref{app-C}).

\section{Harmonic approximation of the intramolecular interactions}
\label{app-B}

\begin{figure*}[t!]
\begin{centering}
\includegraphics[width=140mm]{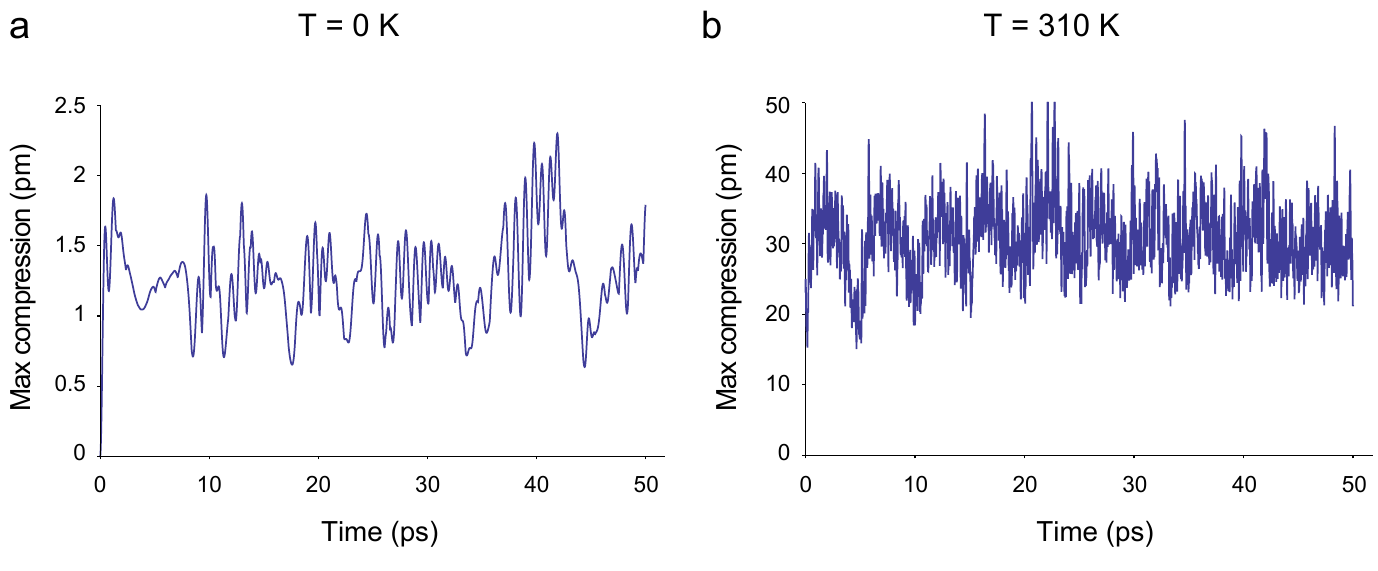}
\par\end{centering}

\caption{\label{fig:9}Dynamics of the maximal compression $\max\left[b_{n,\alpha}-b_{n+1,\alpha}\right]$
for a symmetric soliton in the three spine model with lateral coupling $w_{2}=2w_{1}$ either at zero
temperature, $T=0$~K, or at physiological temperature, $T=310$~K.}
\end{figure*}

Longitudinal and lateral interactions between peptide groups inside
the protein $\alpha$-helix, respectively due to hydrogen bonds or
covalent C-C and C-N bonds, are modeled as harmonic oscillators. This
leads to the possibility that under extreme compression, neighboring
peptide groups are allowed to pass unphysically through each other.
The physical observable $b_{n,\alpha}-b_{n+1,\alpha}<0$ indicates
dilation, whereas $b_{n,\alpha}-b_{n+1,\alpha}>0$ indicates compression
of the lattice between sites $n,\alpha$ and $n+1,\alpha$. Therefore,
to investigate whether the harmonic approximation is justified for
Davydov's model, we have computed the maximal compression
$\max\left[b_{n,\alpha}-b_{n+1,\alpha}\right]$ in the protein as a function of time~$t$.
Because the peptide groups are separated by a distance of 450~pm along the protein $\alpha$-helix spines, the unphysical situation
that peptide groups pass through each other will occur when $b_{n,\alpha}-b_{n+1,\alpha}>450$~pm.
Previously, we have shown that protein solitons correspond to
compression waves in the lattice \cite{Georgiev2019a}. Here, we find
that the maximal compression of the protein lattice occurs when the
soliton bounces off the protein $\alpha$-helix end, and that the compression
is less than $2.5$ pm for a symmetric soliton in the three spine model with lateral coupling $w_{2}=2w_{1}$ at zero
temperature, $T=0$~K (Fig.~\ref{fig:9}a). The introduction of thermal noise
at physiological temperature, $T=310$~K, increases the lattice compression
by $\approx20$ times, but at no time point in the simulation the
compression exceeded 50~pm (Fig.~\ref{fig:9}b). These results demonstrate that
the harmonic approximations for longitudinal and lateral interactions
in the Davydov model are indeed fully justified and no peptide groups
pass unphysically through each other.

\section{Soliton speed}
\label{app-C}

\begin{figure*}[t!]
\begin{centering}
\includegraphics[width=140mm]{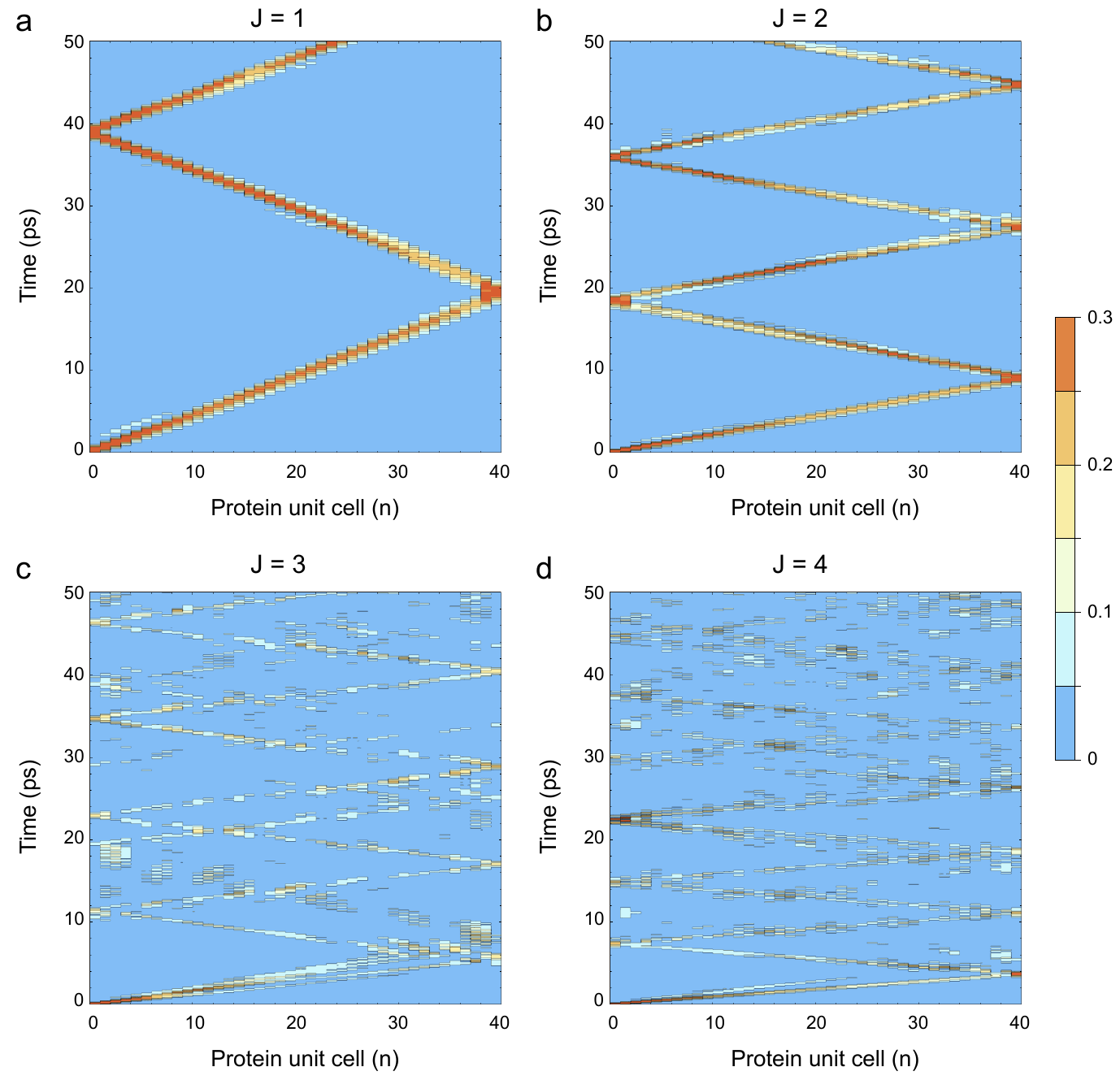}
\par\end{centering}

\caption{\label{fig:10}Dynamics of a soliton generated by unit pulse applied at a single peptide group inside the massive single-chain model of a protein
$\alpha$-helix with $\tilde{M}=3M$ and $\tilde{w}=3w$ at $T=0$~K,
for $Q=3$ amide~I energy quanta and different values of the nearest neighbor dipole-dipole coupling energy $J$ (measured in units of 967.4~$\mu$eV), visualized through the exciton
quantum probability $|a_{n}|^{2}$ at each protein unit cell $n$.
The unit pulse of amide~I energy is applied at the N-end of a
protein $\alpha$-helix with $n_{\max}=40$ unit cells, each of which
is comprised of three peptide groups.}
\end{figure*}

Solitons are generated by pulses of amide~I energy. Due to the discreteness
of the protein lattice, these energy pulses can only be discrete distributions
over the peptide groups. Previously, we have demonstrated that
the soliton speed is inversely related to the soliton's width, namely,
narrow solitons are faster than wider solitons \cite{Georgiev2019a}.
This means that by collecting the total unit probability onto a single
peptide group in the form of a bare amide~I exciton, the discrete
Davydov model will generate the fastest soliton whose speed is
the maximal speed supported by the protein lattice for fixed
nearest neighbor dipole-dipole coupling energy~$J$, spring constant
of the hydrogen bonds~$w$, and strength of the exciton-phonon coupling~$\chi$.
For the massive single-chain model of a protein $\alpha$-helix,
with $J=967.4$~$\mu$eV, $\tilde{w}=3\times 13$~N/m, $\chi_{r}=\chi_{l}=35$~pN, at zero temperature $T=0$~K, the soliton moves at speed $v=947$~m/s (Fig.~\ref{fig:10}a), which is $\frac{1}{4}$ of the speed of sound in the protein lattice
$v_{s}=r\sqrt{\frac{w}{M}}=3722$ m/s.

Previous analysis based on the continuum approximation of Davydov's
equations has found supersonic solutions for certain conditions imposed
on the parameters $J,w,\chi$ \cite{Zolotaryuk1995}. To test whether
the discrete protein lattice can support supersonic solitons, we have
performed simulations by increasing the value of the dipole-dipole
coupling energy $J$, while keeping $w$ and $\chi$ fixed to their established
biological values. By doubling the dipole-dipole coupling energy to
$J=2\times 967.4$~$\mu$eV, the soliton speed increased to 2000 m/s (Fig.~\ref{fig:10}b). Further
increment of the dipole-dipole coupling to $J=3\times 967.4$~$\mu$eV led to soliton dispersal
even though the simulation is performed at zero temperature (Fig.~\ref{fig:10}c).
If one uses the 6~ps time point at which part of the amide~I
energy distribution hits the end of the protein $\alpha$-helix, subsonic
soliton speed of $v=3000$ m/s could be measured, but this speed is
not meaningful because the soliton is rapidly dispersed. Similarly,
if the dipole-dipole coupling is increased to $J=4\times 967.4$~$\mu$eV, the amide~I energy
distribution may hit the end of the protein $\alpha$-helix at 4 ps
time point, registering supersonic speed $v=4500$ m/s, but again this is
not biologically useful as the soliton is highly unstable and easily
dispersed (Fig.~\ref{fig:10}d). From these simulations can be concluded that the
discreteness of the protein lattice has pronounced negative effects
on fast moving solitons, but only minimal effects on slow moving solitons
where the resulting dynamics is comparable with the one obtained from
the continuum approximation.

\section{Multi-exciton ansatz}
\label{app-D}

The choice of ansatz for the multi-exciton quantum state is based
on restricted bosonic Hartree--Fock (RBHF) trial wave function \eqref{eq:Hartree}, which contains
a product of identical single-particle wave functions \cite{Cederbaum2003,Romanovsky2004,Romanovsky2006,Heimsoth2010}.
In the theory of dilute Bose gases, the choice of this restricted ansatz is known as Gross--Pitaevskii approximation \cite{Pitaevskii2003,Sakaguchi2005}.
The restricted ansatz has been introduced by Zolotaryuk \cite{Zolotaryuk1988}
to study multi-particle Davydov solitons in proteins, and is also common in the
study of multiquantal soliton propagation on an optical fiber \cite{Lai1989}.
Expansion of the operator product in \eqref{eq:Hartree} can be achieved
with the multinomial theorem \cite{Georgiev2020a}
\begin{align}
\left(\sum_{n}a_{n}\hat{a}_{n}^{\dagger}\right)^{Q} & =\underset{Q\textrm{ terms}}{\underbrace{\left(a_{1}\hat{a}_{1}^{\dagger}+a_{2}\hat{a}_{2}^{\dagger}+\ldots\right)\left(a_{1}\hat{a}_{1}^{\dagger}+a_{2}\hat{a}_{2}^{\dagger}+\ldots\right)\ldots\left(a_{1}\hat{a}_{1}^{\dagger}+a_{2}\hat{a}_{2}^{\dagger}+\ldots\right)}}\nonumber \\
 & =\underset{Q-1\textrm{ terms}}{\underbrace{\left[a_{1}^{2}\left(\hat{a}_{1}^{\dagger}\right)^{2}+2a_{1}a_{2}\hat{a}_{1}^{\dagger}\hat{a}_{2}^{\dagger}+a_{2}^{2}\left(\hat{a}_{2}^{\dagger}\right)^{2}+\ldots\right]\ldots\left(a_{1}\hat{a}_{1}^{\dagger}+a_{2}\hat{a}_{2}^{\dagger}+\ldots\right)}}\nonumber \\
 & =\sum_{k_{1}+k_{2}+\ldots+k_{n}=Q}\frac{Q!}{k_{1}!k_{2}!\ldots k_{n}!}\left(a_{1}\hat{a}_{1}^{\dagger}\right)^{k_{1}}\left(a_{2}\hat{a}_{2}^{\dagger}\right)^{k_{2}}\ldots\left(a_{n}\hat{a}_{n}^{\dagger}\right)^{k_{n}}
\end{align}
Its main advantage for computational studies is that for a lattice with total of $N$ sites, it requires consideration
of only $N$ differential equations for the $N$ different exciton quantum probability amplitudes.

More general number state method of analysis for \mbox{$Q\geq2$} was proposed by Scott and collaborators \cite{Scott1994} who introduced unrestricted bosonic Hartree--Fock (UBHF) trial wave function, which contains a product of different
single-particle wave functions \cite{Alon2007,Romanovsky2004,Romanovsky2006,Heimsoth2010,Lode2020}
\begin{align}
\prod_{q=1}^{Q}\left(\sum_{n}a_{n,q}\hat{a}_{n}^{\dagger}\right) & =\underset{Q\textrm{ terms}}{\underbrace{\left(a_{1,1}\hat{a}_{1}^{\dagger}+a_{2,1}\hat{a}_{2}^{\dagger}+\ldots\right)\left(a_{1,2}\hat{a}_{1}^{\dagger}+a_{2,2}\hat{a}_{2}^{\dagger}+\ldots\right)\ldots\left(a_{1,Q}\hat{a}_{1}^{\dagger}+a_{2,Q}\hat{a}_{2}^{\dagger}+\ldots\right)}}\nonumber \\
 & =\underset{Q-1\textrm{ terms}}{\underbrace{\left[a_{1,1}a_{1,2}\left(\hat{a}_{1}^{\dagger}\right)^{2}+\left(a_{1,1}a_{2,2}+a_{1,2}a_{2,1}\right)\hat{a}_{1}^{\dagger}\hat{a}_{2}^{\dagger}+a_{2,1}a_{2,2}\left(\hat{a}_{2}^{\dagger}\right)^{2}+\ldots\right]\ldots\left(a_{1,Q}\hat{a}_{1}^{\dagger}+\ldots\right)}}\nonumber \\
 & =\sum_{k_{1}+k_{2}+\ldots+k_{n}=Q}\mathcal{S}\left[a_{1,q}^{k_{1}}a_{2,q}^{k_{2}}\ldots a_{n,q}^{k_{n}}\right]\left(\hat{a}_{1}^{\dagger}\right)^{k_{1}}\left(\hat{a}_{2}^{\dagger}\right)^{k_{2}}\ldots\left(\hat{a}_{n}^{\dagger}\right)^{k_{n}}
\end{align}
where $\mathcal{S}$ is the symmetrization operator \cite{Cederbaum2004,Alon2005,Alon2007} whose action is
to produce a symmetric sum of $\frac{Q!}{k_{1}!k_{2}!\ldots k_{n}!}$ terms taking care
of site repetitions, for example
\begin{equation}
\mathcal{S}\left[a_{1,q}^{1}a_{2,q}^{2}\right]=a_{1,1}a_{2,2}a_{2,3}+a_{1,2}a_{2,1}a_{2,3}+a_{1,3}a_{2,1}a_{2,2}
\end{equation}
In total, there will be $N$ multichoose $Q$ different symmetrized
quantum probability amplitudes
\begin{equation}
\left(\left(\begin{array}{c}
N\\
Q
\end{array}\right)\right)=\left(\begin{array}{c}
N+Q-1\\
Q
\end{array}\right)=\frac{(N+Q-1)!}{Q!(N-1)!}
\end{equation}
This means that with the unrestricted ansatz, the number of differential
equations rapidly grows with $Q$.
In the case of single-chain lattice with $N=40$ sites, one would need a supercomputer in order to solve
the large number of differential equations: 820 for $Q=2$, 11480
for $Q=3$, and 123410 for $Q=4$. Simulations of the three spine model with $N=120$ sites would be even more demanding
on resources as there will be 295240 equations for $Q=3$.
Thus, the unrestricted ansatz makes it practically impossible to explore computationally the three-spine model.
In contrast, the utility of the restricted ansatz \eqref{eq:Hartree} is that it provides reasonable means
for computer simulations of Davydov's three-spine model.

Investigating the numerical accuracy of the restricted exciton ansatz \eqref{eq:Hartree}, though an interesting open question, 
it is however, beyond the scope of this present work. Here, we have focused our efforts on elucidating how different modifications
of the model Hamiltonian may affect the thermal stability of the generated solitons.
Through systematic exploration of the available parameter space, we have been able to optimize the
initial conditions and obtain thermally stable solitons. Our results
challenge previous claims that the system of differential equations
derived by the restricted exciton ansatz~\eqref{eq:Hartree} does not possess sufficient thermal
stability \cite{Kerr1990,Lomdahl1990}. We have shown that such claims
originate from improper consideration of a single protein $\alpha$-helix
spine in isolation. If the contribution of the other two lateral spines
is properly accounted for, either by tripling the mass of each lattice
site in the massive single-chain model of a protein $\alpha$-helix or by
inclusion of lateral coupling in the three spine model,
the soliton lifetime exceeds 30~ps at physiological temperature, $T=310$~K,
and is sufficient for propagation over biologically meaningful distance.

\section{Dynamics of discrete sech-squared solitons in the massive single-chain model of a protein \texorpdfstring{$\alpha$}{alpha}-helix}
\label{app-E}

Discretization of continuous distributions into the form given by Eq.~\eqref{eq:pulse}
was performed by centering the exciton energy pulse onto 5 peptide
groups, and controlling the spread so that 95\% of the exciton probability
is localized in the interval from $-2.5r$ to $2.5r$. Then, definite
integrals were computed with a bin size of $r=0.45$ nm, which is
the distance between neighboring peptide groups along the spine. The resulting probabilities $\left|A_{n}\right|^{2}$ were
truncated to 3 decimal places and their total sum was normalized to~1.

\begin{figure*}[t!]
\begin{centering}
\includegraphics[width=140mm]{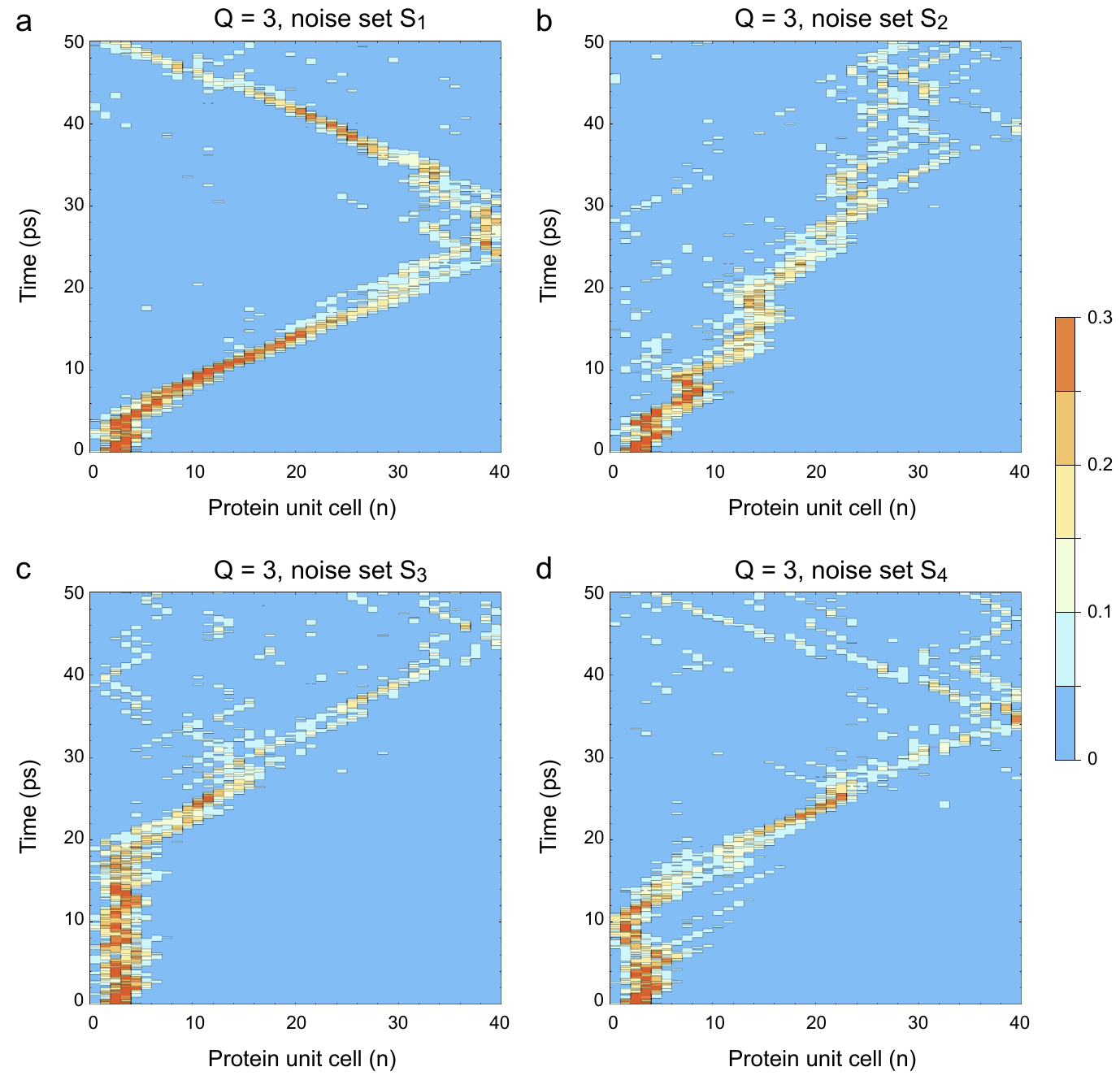}
\par\end{centering}

\caption{\label{fig:11}Dynamics of a sech-squared soliton moving inside the massive single-chain model of a protein
$\alpha$-helix with $\tilde{M}=3M$ and $\tilde{w}=3w$ at $T=310$~K,
for $Q=3$ amide~I energy quanta under the influence of different
sets $S_{i}$ of Wiener processes, visualized through the exciton
quantum probability $|a_{n}|^{2}$ at each protein unit cell $n$.
The sech-squared pulse of amide~I energy is applied at the N-end of a
protein $\alpha$-helix with $n_{\max}=40$ unit cells, each of which
is comprised of three peptide groups. The protein soliton starts from
the same initial thermalized state of the lattice but exhibits different
trajectories for different actualizations $S_{i}$ of the thermal noise.}
\end{figure*}

For the Gaussian pulse, we have used $\sigma=1.2755\,r$ to compute
the bins
\begin{equation}
\left|A_{n}\right|^{2}=\frac{100}{95}\times\int_{(n-0.5)r}^{(n+0.5)r}\frac{1}{\sqrt{2\pi}\sigma}\exp\left(-\frac{x^{2}}{2\sigma^{2}}\right)
\end{equation}
with corresponding moduli $A_{0}=\sqrt{0.322}$, $A_{1}=\sqrt{0.24}$,
$A_{2}=\sqrt{0.099}$.

To test the possible effect of initial sech-squared shape of the exciton
pulse on thermal stability, we have used $\sigma=1.3647\,r$ to compute
the bins 
\begin{equation}
\left|A_{n}\right|^{2}=\frac{100}{95}\times\int_{(n-0.5)r}^{(n+0.5)r}\frac{1}{2\sigma}\textrm{sech}^{2}\left(\frac{x}{\sigma}\right)
\end{equation}
with corresponding moduli $A_{0}=\sqrt{0.37}$, $A_{1}=\sqrt{0.237}$,
$A_{2}=\sqrt{0.078}$.

Computer simulations have shown that the dynamics of the discrete
sech-squared exciton pulse (Fig.~\ref{fig:11}) is almost identical to the dynamics
produced with discrete Gaussian pulse (Fig.~\ref{fig:7}). Thus, the thermal stability of discrete
sech-squared and discrete Gaussian pulses is the same. This indicates that
the effects of discretization of the protein lattice dominate over
any differences in the initial shape of continous distributions of
free energy that are released from ATP.

\section{Relationship between the massive single-chain model and the three-spine model}
\label{app-F}

At zero temperature, $T=0$ K, symmetric initial distributions of the exciton energy in the three-spine
model lead to quantum dynamics that is mathematically equivalent to
the one obtained with the massive single-chain model of a protein $\alpha$-helix,
with $\tilde{M}=3M$ and $\tilde{w}=3w$. To show that, we explicitly
write the implications from the initial symmetry as
\begin{align}
a_{n,\alpha} & =a_{n,\alpha\pm1}=\frac{1}{\sqrt{3}}\bar{a}_{n}\label{eq:sym-1}\\
b_{n,\alpha} & =b_{n,\alpha\pm1}=b_{n}\label{eq:sym-2}
\end{align}
Substitution of \eqref{eq:sym-1} and \eqref{eq:sym-2} in the three-spine
model \eqref{eq:full-1} gives
\begin{equation}
\imath\hbar\frac{d}{dt}\frac{1}{\sqrt{3}}\bar{a}_{n}=-J_{1}\frac{1}{\sqrt{3}}\left(\bar{a}_{n+1}+\bar{a}_{n-1}\right)+J_{2}\frac{1}{\sqrt{3}} 2\bar{a}_{n}+\left[\chi_{r}b_{n+1}+(\chi_{l}-\chi_{r})b_{n}-\chi_{l}b_{n-1}\right]\frac{1}{\sqrt{3}}\bar{a}_{n}
\end{equation}
The common factor of $\frac{1}{\sqrt{3}}$ can be canceled out. After introducing
the gauge transformation $\bar{a}_{n}=a_{n}e^{-\frac{\imath}{\hbar}2J_{2}t}$,
we obtain
\begin{equation}
\imath\hbar\frac{d}{dt}a_{n}=-J_{1}\left(a_{n+1}+a_{n-1}\right)+\left[\chi_{r}b_{n+1}+(\chi_{l}-\chi_{r})b_{n}-\chi_{l}b_{n-1}\right]a_{n}\label{eq:sym-3}
\end{equation}
This is exactly \eqref{eq:gauge-1} from the massive single-chain model.

At zero temperature, $T=0$ K, we do not need to introduce any Langevin
terms for the phonon lattice. Substitution of \eqref{eq:sym-1} and
\eqref{eq:sym-2} in the three-spine model \eqref{eq:full-2} further gives
\begin{equation}
\frac{d^{2}}{dt^{2}}b_{n}=\frac{w_{1}}{M}\left(b_{n-1}-2b_{n}+b_{n+1}\right)-\frac{Q}{M}\frac{1}{3}\left[\chi_{r}\left|a_{n-1}\right|^{2}+(\chi_{l}-\chi_{r})\left|a_{n}\right|^{2}-\chi_{l}\left|a_{n+1}\right|^{2}\right]\label{eq:sym-4}
\end{equation}
This is exactly \eqref{eq:gauge-2} as seen by taking into account that $\frac{w_{1}}{M}=\frac{\tilde{w}}{\tilde{M}}$
and $\frac{Q}{M}\frac{1}{3}=\frac{Q}{\tilde{M}}$. The lateral coupling
term is eliminated due to $b_{n,\alpha-1}-2b_{n,\alpha}+b_{n,\alpha+1}=0$.
Noteworthy, the origin of the factor of $\frac{1}{3}$ comes from
the equal distribution of the exciton amplitudes across the three
spines, and not from triple mass of the lattice.

It is exactly due to the mathematical equivalence between the system
\eqref{eq:sym-3}, \eqref{eq:sym-4} and the system \eqref{eq:gauge-1}, \eqref{eq:gauge-2}, that
we can interpret the massive single-chain model of a protein $\alpha$-helix
in two valid ways:

If we have $\sum_{n}\left|a_{n}\right|^{2}=1$, we can state that
the massive single-chain model of a protein $\alpha$-helix is a coarse-grained
description of three-spine model with symmetric initial distribution
of the exciton energy across the three spines, where the coarse graining
comes from combining the probabilities within each protein unit cell
together
\begin{equation}
P(a_{n})=\sum_{\alpha}\left|a_{n,\alpha}\right|^{2}=3\times\frac{1}{3}\left|a_{n}\right|^{2}=\left|a_{n}\right|^{2}
\end{equation}

Alternatively, if we have $\sum_{n}\left|a_{n}\right|^{2}=\frac{1}{3}$,
we can state that the system \eqref{eq:gauge-1}, \eqref{eq:gauge-2} describes not a massive single-chain
model of protein $\alpha$-helix, but one of three identical spines
in the three-spine model with symmetric initial conditions.

Because the latter two interpretations involve only an overall scaling
factor of $\frac{1}{3}$, it is clear that the observed soliton dynamics
in the plots remains unchanged. This also explains why the model of
single isolated spine with $Q=1$ is equivalent to symmetric three-spine
model with $Q=3$ at zero temperature.

In the presence of thermal noise at physiological temperature $T=310$~K, the above derivations may not be always justified because the thermal
noise injected at each of the three spines is statistically independent.
This means that in the absense of lateral coupling between the spines,
$w_{2}=0$, the presence of thermal noise at each spine will immediately
destroy the symmetry in the phonon displacements, hence $b_{n,\alpha-1}-2b_{n,\alpha}+b_{n,\alpha+1}\neq0$.
In the limit of infinitely stiff lateral coupling, $w_{2}\to\infty$,
however, the protein unit cell will behave as a single block with
triple mass, $\tilde{M}=3M$, for which the phonon displacements are
fixed laterally by \eqref{eq:sym-2}. Since the contribution by the lateral coupling term
is eliminated due to $b_{n,\alpha-1}-2b_{n,\alpha}+b_{n,\alpha+1}=0$,
the massive single-chain model of a protein $\alpha$-helix could be
viewed as providing a coarse-grained description of three-spine model
with $w_{2}\to\infty$ under symmetric initial conditions.

\section{Dynamics of asymmetric sech-squared solitons in the three-spine model}
\label{app-G}

\begin{figure*}[t!]
\begin{centering}
\includegraphics[width=140mm]{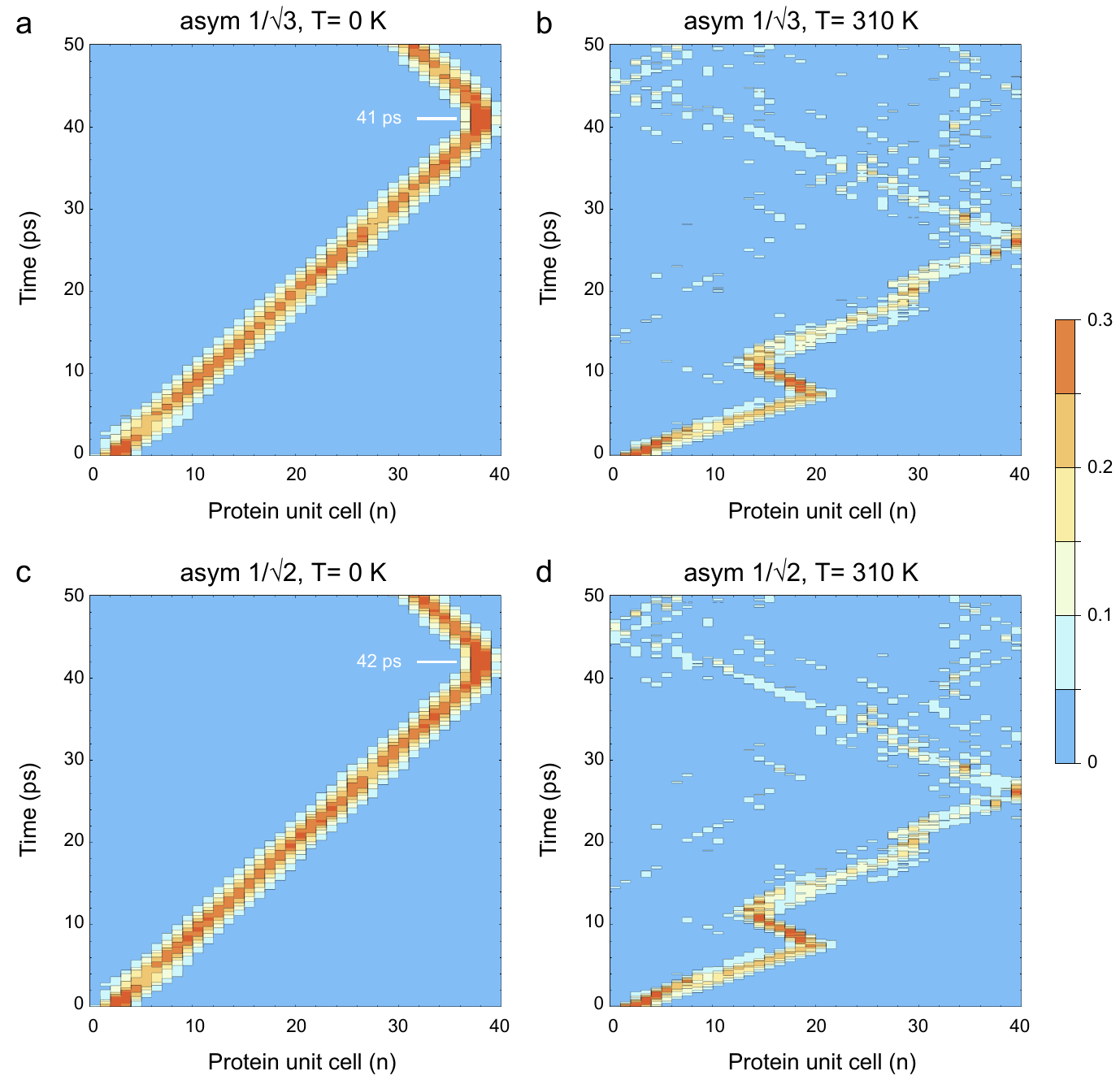}
\par\end{centering}

\caption{\label{fig:12}Dynamics of asymmetric solitons moving inside the three spine model of
a protein $\alpha$-helix at zero temperature, $T=0$~K, or in the presence of thermal noise, $T=310$~K,
for $Q=3$ amide~I energy quanta, visualized through the exciton
quantum probability $\sum_\alpha |a_{n,\alpha}|^{2}$ at each protein unit cell $n$.
The initial energy pulse is given by \eqref{eq:pulse-3a} for panels (a,b) and \eqref{eq:pulse-3b} for panels (c,d).
The fixed set of 120 Wiener processes for the simulations with thermal noise (b,d) is the same set used in Fig.~\ref{fig:8}.}
\end{figure*}

Previous investigations of the three-spine model within the continuum approximation, have revealed several types of solitons, among which the ground state soliton (entangled or hybrid soliton) whose energy is 50~times lower than the energy of a soliton in a single spine, and, thus, expected to have much higher thermal stability than the latter one \cite{Brizhik2004}. For the totally symmetric soliton, the three spines are excited with the same phase, $\Delta\omega_2 = 0$, whereas for hybrid solitons the excitations in the spines have lateral phase shifts, $\Delta\omega_2 = \pm \frac{2\pi}{3}$.

Because the parameter space of the three-spine model is vastly increased compared to the massive single-chain model, a thorough numerical investigation would require its own dedicated study. Nevertheless, here we present numerical simulations for the three-spine model with lateral coupling $w_2=2 w_1$ in the presence of $Q=3$ exciton quanta under two types of asymmetric initial conditions based on the discrete sech-squared exciton pulse: (i) with lateral asymmetry in the phase and equal amplitude of the exciton pulse on all three spines, or (ii) with lateral asymmetry in the phase and equal amplitude of the exciton pulse on only two of the three spines.

The first type of asymmetric soliton (Fig.~\ref{fig:12}a,b) was excited with the initial pulse
\begin{equation}
\frac{1}{\sqrt{3}}e^{\imath\alpha\Delta\omega_{2}}\left\{ A_{2,\alpha}e^{-\imath2\Delta\omega_{1}},A_{1,\alpha}e^{-\imath\Delta\omega_{1}},A_{0,\alpha},A_{1,\alpha}e^{+\imath\Delta\omega_{1}},A_{2,\alpha}e^{+\imath2\Delta\omega_{1}}\right\} 
\label{eq:pulse-3a}
\end{equation}
where $A_{0,\alpha}=\sqrt{0.37}$, $A_{1,\alpha}=\sqrt{0.237}$, $A_{2,\alpha}=\sqrt{0.078}$,
and the phase modulation was present both along the spines $\Delta\omega_1=\frac{\pi}{12}$ and laterally between the spines $\Delta\omega_2=\frac{2\pi}{3}$.

The second type of asymmetric soliton (Fig.~\ref{fig:12}c,d) was excited with the initial pulse
\begin{equation}
\frac{1}{\sqrt{2}}\textrm{sign}(\alpha)e^{\imath\alpha\Delta\omega_{2}}\left\{ A_{2,\alpha}e^{-\imath2\Delta\omega_{1}},A_{1,\alpha}e^{-\imath\Delta\omega_{1}},A_{0,\alpha},A_{1,\alpha}e^{+\imath\Delta\omega_{1}},A_{2,\alpha}e^{+\imath2\Delta\omega_{1}}\right\} 
\label{eq:pulse-3b}
\end{equation}
where the spine indexed with $\alpha=0$ was empty and the exciton energy was distributed onto the other two spines indexed with $\alpha=1,2$ for $A_{0,\alpha}=\sqrt{0.37}$, $A_{1,\alpha}=\sqrt{0.237}$, $A_{2,\alpha}=\sqrt{0.078}$,  $\Delta\omega_1=\frac{\pi}{12}$ and $\Delta\omega_2=\frac{2\pi}{3}$.

At zero temperature, $T=0$~K, the first type of asymmetric soliton that is distributed over all three spines moves at speed of 439 m/s (Fig.~\ref{fig:12}a), 
whereas the second type of asymmetric soliton that is distributed only over two of the three spines is a little bit slower, due to stronger self-trapping, and moves at speed of 429 m/s  (Fig.~\ref{fig:12}c).
Interestingly, the thermal stability of both types of asymmetric solitons is about 30~ps (Fig.~\ref{fig:12}b,d), and virtually identical to the symmetric soliton (Fig.~\ref{fig:8}c).
This lack of significant difference in thermal stability between symmetric and asymmetric solitons could be very useful from a biological perspective, namely, the discrete lattice of peptide groups in protein $\alpha$-helices could utilize any initial distribution of metabolic free energy with the same high efficiency. Or to put it differently, our numerical results suggest that the proteins do not require some highly specialized initial state in order to generate solitons. Instead, any sufficiently localized energy pulse, regardless of its initial distribution onto the spines could reshape itself into the proper soliton-like form that guarantees thermal stability for at least 30~ps.

\end{document}